\documentclass[aps,prl,twocolumn,preprintnumbers,amsmath,amssymb,superscriptaddress]{revtex4-2}
\usepackage{amsmath,graphicx}
\usepackage[utf8]{inputenc}
\usepackage[T1]{fontenc}
\usepackage{xcolor}
\usepackage{textcomp}
\usepackage{bm}

\usepackage{siunitx}
\usepackage{physics}
\usepackage{amsmath}
\usepackage{tikz}
\usepackage{mathdots}
\usepackage{yhmath}
\usepackage{cancel}
\usepackage{color}
\usepackage{array}
\usepackage{multirow}
\usepackage{amssymb}
\usepackage{gensymb}
\usepackage{tabularx}
\usepackage{extarrows}
\usepackage{booktabs}
\usetikzlibrary{fadings}
\usetikzlibrary{patterns}
\usetikzlibrary{shadows.blur}
\usetikzlibrary{shapes}

\usepackage{color}
\definecolor{LinkColor}{rgb}{0.75,0.0,0.2}

\usepackage{hyperref}
\hypersetup{
	pdfauthor={good guys},
	pdftitle={good title},
	colorlinks=true,
	citecolor=LinkColor,
	linkcolor=LinkColor,
	urlcolor=LinkColor,
}

\usepackage{listings}
\definecolor{lightgray}{gray}{1}

\begin{document}
\title{Entanglement Structure and Information Protection in Noisy Hybrid Quantum Circuits}

\author{Shuo Liu}
\affiliation{Institute for Advanced Study, Tsinghua University, Beijing 100084, China}
\author{Ming-Rui Li}
\affiliation{Institute for Advanced Study, Tsinghua University, Beijing 100084, China}
\author{Shi-Xin Zhang}
\email{shixinzhang@tencent.com}
\affiliation{Tencent Quantum Laboratory, Tencent, Shenzhen, Guangdong 518057, China}
\author{Shao-Kai Jian}
\email{sjian@tulane.edu}
\affiliation{Department of Physics and Engineering Physics, Tulane University, New Orleans, Louisiana, 70118, USA}

\date{\today}

\begin{abstract}
In the context of measurement-induced entanglement phase transitions, the influence of quantum noises, which are inherent in real physical systems, is of great importance and experimental relevance. 
In this Letter, we present a comprehensive theoretical analysis of the effects of both temporally uncorrelated and correlated quantum noises on entanglement generation and information protection. 
This investigation reveals that entanglement within the system follows $q^{-1/3}$ scaling for both types of quantum noises, where $q$ represents the noise probability. 
The scaling arises from the Kardar-Parisi-Zhang fluctuation with effective length scale $L_{\text{eff}} \sim q^{-1}$. 
More importantly, the information protection timescales of the steady states are explored and shown to follow $q^{-1/2}$ and $q^{-2/3}$ scaling for temporally uncorrelated and correlated noises, respectively. 
The former scaling can be interpreted as a Hayden-Preskill protocol, while the latter is a direct consequence of Kardar-Parisi-Zhang fluctuations. 
We conduct extensive numerical simulations using stabilizer formalism to support the theoretical understanding. 
This Letter not only contributes to a deeper understanding of the interplay between quantum noises and measurement-induced phase transition but also provides a new perspective to understand the effects of Markovian and non-Markovian noises on quantum computation.
\end{abstract}

\maketitle

\textit{Introduction.---}
The competition between unitary evolution and non-unitary monitored measurements gives rise to a dynamical phase transition known as the measurement-induced phase transition (MIPT)~\cite{PhysRevB.98.205136, PhysRevB.100.134306, PhysRevX.9.031009, PhysRevLett.126.060501, PRXQuantum.2.040319, PhysRevX.12.011045, PhysRevX.10.041020, PhysRevLett.125.030505, PhysRevB.99.224307, PhysRevB.100.064204, PhysRevB.101.104301, PhysRevB.103.174309, PhysRevB.103.104306, PhysRevB.101.104302, PhysRevB.107.214203, hokeMeasurementinducedEntanglementTeleportation2023, PhysRevB.105.104306, PhysRevB.102.014315, PhysRevB.106.214316, PhysRevB.106.144311, PRXQuantum.4.030333}.
The entanglement within the system undergoes a transition from a volume-law phase to an area-law phase as the measurement probability increases. 
Theoretical understanding of MIPT~\cite{PhysRevB.101.104301, PhysRevB.101.104302} reveals its connection to the order-disorder transition of a classical spin model through the mapping between the hybrid quantum circuit and an effective statistical model. 
Building upon this theoretical understanding, MIPT has been extensively investigated in various systems~\cite{PhysRevX.12.041002, PhysRevB.107.014308, PhysRevLett.129.120604, han2023entanglement, PhysRevB.108.054307, PhysRevLett.131.020401, PhysRevLett.127.140601, PhysRevB.106.224305, MIPT_SYK_2, Biella_2021, PhysRevB.103.224210, PhysRevB.105.L241114, NonlocalMIPT_Qi,PRXQuantum.2.010352, NonlocalMIPT_Quantum, PhysRevB.104.094304, PhysRevLett.128.010605, PhysRevLett.128.010604, PhysRevResearch.4.013174, PhysRevLett.128.010603, 10.21468/SciPostPhysCore.5.2.023, Zhang2022universal, PhysRevB.108.L041103, ravindranath2023free, o2022entanglement, PhysRevLett.130.120402, kelly2024generalizing, kelly2023information, 10.21468/SciPostPhysCore.7.1.011, PhysRevB.108.075151, doggen2023ancilla}.

The entanglement structure and information protection are intricately connected~\cite{PhysRevLett.125.030505, PhysRevLett.111.127205, PhysRevX.7.031016, PhysRevX.11.011030, QI_NP, PhysRevX.10.041020}. 
For a noiseless monitored random circuit in a volume law phase (also dubbed as error-resilient phase), the encoded information will remain in the system for an infinitely long time. 
However, in real experiments, inevitable quantum noise from the environment can disrupt entanglement within the system. \cite{hokeMeasurementinducedEntanglementTeleportation2023}. From the entanglement perspective, it has been demonstrated that quantum noise can be treated as a symmetry-breaking field in the effective statistical model, resulting in a single area-law entanglement phase and the disappearance of MIPT with infinitesimal noise strength $q$~\cite{BAO2021168618, Noise_bulk, jian2021quantum, PhysRevB.108.104310, PhysRevB.107.014307, PhysRevB.108.104310, PhysRevB.107.L201113}. Nevertheless, investigations on the information protection capacity in noisy hybrid quantum circuits are rare and strongly required, especially necessitating the identification of a characteristic timescale for information protection,
because noise errors are more general and common than measurement errors on quantum devices for quantum error correction.

While temporally correlated measurements in MIPT have been explored~\cite{PhysRevB.107.L220204, PhysRevB.108.184204}, understanding of the distinction and connection between temporally uncorrelated and correlated quantum noises in MIPT setups remain elusive. In this Letter, we investigate quantum noises with distinct temporal correlations in monitored circuits from both entanglement and information perspectives and primarily concentrate on the latter. The temporally uncorrelated noise can be regarded as the Markovian limit, and the correlated noises correspond to the strong non-Markovian limit~\cite{Breuer2016_z}. 
We not only provide a thorough theoretical understanding of the same area law $q^{-1/3}$ scaling from the entanglement perspective in the original volume law phase of MIPT but also propose a steady state information protection setup where the timescales of information protection reveal the distinctions between the effects of temporally correlated and uncorrelated quantum noises. The theoretical prediction for this timescale is crucial for a better understanding of the effects of quantum noises on information protection and is potentially relevant for quantum error corrections and quantum error mitigations \cite{Cai2022_z, Temme2017_b,  Kim2021b_z, Zhang2021d_z, Kim2023_z}. 

To quantify the entanglement of the mixed states generated by the noisy hybrid quantum circuits~\cite{PhysRevA.54.3824, PhysRevLett.80.5239}, 
we use the mutual information~\cite{nielsen2010quantum} between the left and right half chains, defined as $I_{A:B} = S_{A} + S_{B} - S_{AB}$, where $S_{\alpha}$ is the von Neumann entropy of region $\alpha$. 
Compared to the logarithmic entanglement negativity which is a measure of entanglement for mixed states~\cite{PhysRevA.65.032314, PhysRevLett.95.090503, PhysRevLett.109.130502, Calabrese_2013, PhysRevB.99.075157, PhysRevLett.125.116801, PhysRevB.102.235110, 
Negativity_PhysRevLett20_Wu, Negativity_PRXQuantum21, Negativity_Shapourian}, mutual information shows qualitatively similar behaviors and provides a more intuitive understanding within the framework of the statistical model~\cite{PhysRevLett.129.080501, PhysRevB.107.L201113}. In terms of the corresponding effective statistical model, as shown in the Supplemental Materials (SM)~\footnote{See the Supplementary Materials for more details, including (I) the introduction to the effective statistical model, (II) the analytical understanding in the presence of temporally uncorrelated quantum noises, (III) the analytical understanding in the presence of temporally correlated quantum noises, (IV) numerical results for larger measurement rate $p_{m}>p_{m}^{c}$,
(V) information protection with other setups, (VI) connection between information protection and time correlation of quantum noises, (VII) details of numerical simulation.}, $S_{\alpha}$ is expressed as the free energy difference of a classical spin model with specific boundary conditions. 
The presence of temporally correlated noises induces an effective length scale $L_{\text{eff}} \sim q^{-1}$ and the free energy scaling can be analytically obtained from the Kardar-Parisi-Zhang (KPZ) theory directly~\cite{PhysRevLett.56.889, PhysRevLett.55.2923, PhysRevLett.55.2924, PhysRevA.39.3053, PhysRevLett.129.080501, Gueudré_2012,barraquandHalfSpaceStationaryKardar2020} with $L_{\text{eff}}$, leading to $q^{-1/3}$ scaling ~\cite{PhysRevB.107.L201113, Note1}. However, the extension of KPZ understanding becomes challenging in the presence of temporally uncorrelated noises with random space-time locations. We provide analytical solutions for the mechanism of $q^{-1/3}$ scaling for the latter case.

More importantly, although the effects of noises with different temporal correlations are indistinguishable from the entanglement perspective, the information protection timescales of the steady states can reveal the distinctions. With temporally correlated noise, the $q^{-2/3}$ timescale emerges as the average height of the domain wall in the statistical model. 
This scaling is given by KPZ theory with $L_{\text{eff}} \sim q^{-1}$ and the wandering exponent $\chi=2/3$. 
The temporally uncorrelated noise is more subtle as the encoded information itself can modify the domain wall configuration in the statistical model, resulting in the $q^{-1/2}$ scaling for information protection. 
This scaling draws an interesting analogy between the hybrid circuits setup and the Hayden-Preskill protocol for black holes~\cite{PatrickHayden_2007}. 
We also validate the theoretical predictions with extensive numerical results from the large-scale stabilizer circuit simulation. 

\begin{figure}[ht]
\centering
\includegraphics[width=0.44\textwidth, keepaspectratio]{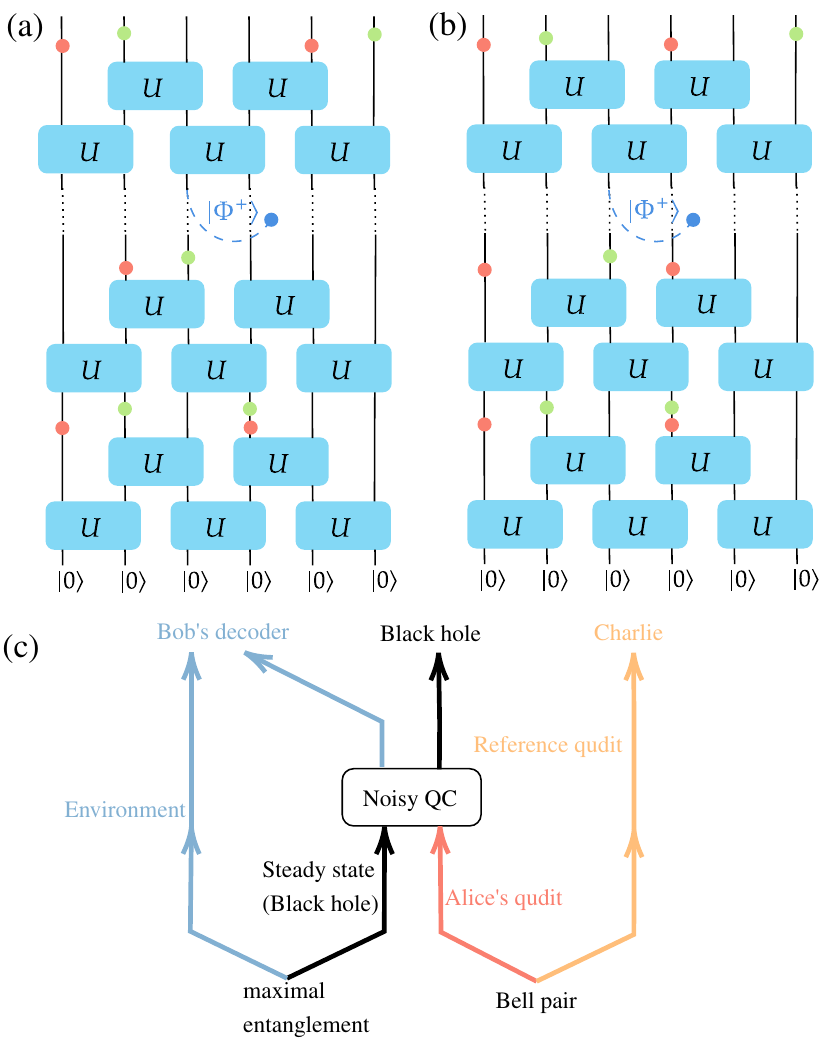}
\caption{Circuit diagram in the presence of temporal uncorrelated (a) and temporal correlated (b) quantum noises (red circles). The initial state is $\vert 0 \rangle^{\otimes L}$ and the projective measurements are represented by green circles. After the system reaches the steady state, a reference qudit (blue circle) is maximally entangled with the middle qudit via forming a Bell pair to encode one-qudit information. (c) shows the corresponding diagram of the Hayden-Preskill protocol. The steady state can be regarded as a black hole and Alice throws one-qudit information into the black hole to destroy it. The timescale of information protection corresponds to the time required for Bob to decode the information from collecting the qudits released by Hawking radiation.}
\label{fig:fig1main}
\end{figure}

\textit{Setup.---} We consider a one-dimensional system with $L$ $d$-qudits under the hybrid evolution with brick-wall random 2-qudit unitary gates in the presence of the projective measurements with probability $p_{m}$ and the quantum noises with probability $q$.  Different quantum channels can be employed to model quantum noise, yielding qualitatively similar results~\cite{Note1}, while unital and non-unital quantum channels induce very different consequences in variational quantum algorithms~\cite{mele2024noise}.
We focus on the region $p_m<p_m^c$, where $p_m^c$ corresponds to the MIPT critical point.
The initial state is chosen as a product state $\vert 0 \rangle^{\otimes L}$ and each 2-qudit gate is independently drawn from the Haar ensemble (or from random 2-qubit Clifford ensemble in numerical simulation). 
The space-time locations of the projective measurements and temporally uncorrelated quantum noises are random as shown in Fig. \ref{fig:fig1main} (a). 
In contrast, only the spatial locations of temporally correlated quantum noises are random, i.e. the locations of quantum noise show a stripe pattern in the time direction as shown in Fig. \ref{fig:fig1main} (b). 
This is the strongest limit of non-Markovianity, where the noise occurrence correlation at any two different time slices at the same spatial position is constant 1.

We calculate the von Neumann entropy $S_{\alpha}$ and mutual information $I_{A:B}$ in the steady states where $I_{A:B}$ saturates and remains constant to reveal entanglement structures. We note that the time required to reach the steady state is $O(q^{-1})$ and size-independent~\footnote{We note that the time required to reach the steady state in the noisy hybrid quantum circuits is $O(q^{-1})$ and size-independent, regardless of the temporal correlation of quantum noises, arising from the competition between the spin configuration where all the spins are fixed to $\mathbb{C}$ with free energy $O(qLt)$ and the spin configuration with domain wall where the leading term of free energy is $O(L)$.}, differing from that of MIPT without noises. To examine the capabilities of information protection, after reaching the steady state, a reference qudit (R) is maximally entangled with the middle qudit of the system by forming a Bell pair to encode one-qudit quantum information as shown in Fig. \ref{fig:fig1main}, see the SM for details~\cite{Note1}. 
Subsequently, we measure the mutual information $I_{AB:R}$ between the system qudits and the reference qudit to study the timescale for information protection.

\textit{Statistical model.---}
We introduce the mapping between the hybrid quantum circuit and the effective statistical model (See the SM~\cite{Note1} for more details).
To calculate the von Neumann entropy $S_{\alpha}$ from the free energy of the effective statistical model, $S_{\alpha}$ is expressed as $ S_{\alpha} = \underset{n \rightarrow 1}{\lim}S^{(n)}_{\alpha} = \underset{n \rightarrow 1}{\lim} \frac{1}{1-n} \mathbb{E}_{\mathcal{U}} \log\frac{\tr \rho_{\alpha}^{n}}{(\tr \rho)^{n}}$, where $\mathbb{E}_{\mathcal{U}}$ represents the average over random two-qudit unitary gates, $\rho_{\alpha}$ is the reduced density matrix of region $\alpha$ and $S^{(n)}_{\alpha}$ is the $n$-th order Renyi entropy given by $ S_{\alpha}^{(n)} = \frac{1}{1-n} \mathbb{E}_{\mathcal{U}} \log \frac{\Tr((C_{\alpha} \otimes I_{\bar{\alpha}}) \rho^{\otimes n})}{\Tr ( (I_{\alpha} \otimes I_{\bar{\alpha}}) \rho^{\otimes n})}$, where $C$ and $I$ are cyclic and identity permutations in $n$-copies replicated Hilbert space, respectively. 
The average of the logarithmic function can be evaluated with the help of the replica trick~\cite{nishimori2001statistical, kardar2007statistical}, $ S_{\alpha} = \lim_{\underset{n \rightarrow 1}{k \rightarrow 0 }} \frac{1}{k(1-n)} \log \left\{ \frac{Z_{\alpha}^{(n,k)}}{Z_{0}^{(n,k)}} \right\} = \lim_{\underset{n \rightarrow 1}{k \rightarrow 0 }} \frac{1}{k(n-1)} \left[ F_{\alpha}^{(n,k)} - F_{0}^{(n,k)}\right]$, where $Z^{(n,k)}$ corresponds to the partition function of a ferromagnetic spin model in the triangular lattice obtained by averaging over the random two-qudit unitary gates. 
We note that $\frac{1}{k(n-1)}F^{(n,k)}$ is independent of $(n,k)$ and thus the limit can be safely taken. 
At each site of the triangular lattice, the degrees of freedom are formed by the permutation-valued spins $\sigma$ defined on the permutation group $S_{nk}$~\cite{Note1}. 
The time runs from the bottom to the top in this effective spin model.
The von Neumann entropy is represented as the free energy difference for the spin model with different top boundary conditions: $\mathbb{C}_{\alpha} \otimes \mathbb{I}_{\bar{\alpha}}$ and $\mathbb{I}_{\alpha} \otimes \mathbb{I}_{\bar{\alpha}}$ for $Z_{\alpha}^{(n,k)}$ and $Z_{0}^{(n,k)}$ respectively, where $\mathbb{C} = C^{\otimes k}$ and $\mathbb{I} = I^{\otimes k}$. The bottom boundary is free due to the initial product state
and therefore $Z^{(n,k)}_{0} = d^{0}$ and $F^{(n,k)}_{0}=0$~\cite{Note1}. In the following discussion, we focus on the most dominant spin configuration in the large $d$ limit, meaning that the partition function $Z^{(n,k)}$ corresponds to the weight of the dominant spin configuration.

In the absence of quantum noises and measurements, $Z^{(n,k)}_{AB}$ is $d^{0}$ because the dominant spin configuration is that all the spins are $\mathbb{C}$. 
Thus, $S_{AB}=0$ consistent with the fact that the steady state is pure. However, a domain wall separating regions $\mathbb{C}$ and $\mathbb{I}$ with unit energy $\vert \mathbb{C} \vert = k(n-1)$ is formed due to the fixed top boundary condition, which is unique because of the unitary constraint~\cite{Uaverage_Qi, PRXQuantum.4.010331}. 
Consequently, $F^{(n,k)}_{\alpha} = \vert \mathbb{C} \vert L_{A}$ and thus $S_{A}=L_{A}$ obeys a volume law.

\begin{figure}[ht]
\centering
\includegraphics[width=0.46\textwidth, keepaspectratio]{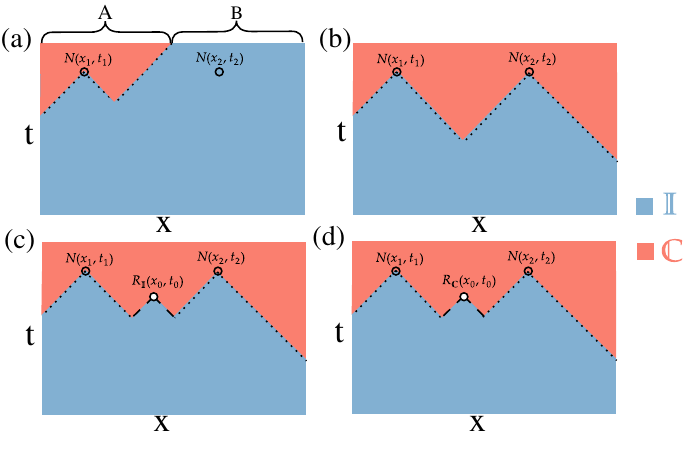}
\caption{The schematic dominant spin configurations: the upper panel is related to entanglement generation while the lower panel is related to information protection setup with extra Bell pair forming at $(x_0, t_0)$. $x$-axis and $y$-axis correspond to spatial and time dimensions respectively. Different colors represent the different classical spin configurations in the statistical model. $N$ and $R$ represent the location for quantum noise and the Bell pair, respectively. Other quantum noises within the $\mathbb{I}$ domain are not shown. (a) shows the dominant spin configuration for $F^{(n,k)}_{A}$. The midpoint on the top boundary will change the length scale of the domain wall near it. (b) shows the dominant spin configuration for $F^{(n,k)}_{AB \cup R}$. (c)(d) show the dominant spin configurations for $F^{(n,k)}_{AB}$ and $F^{(n,k)}_{AB \cup R}$ where spin freedom at $(x_0, t_0)$ are fixed to $\mathbb{I}$ and $\mathbb{C}$ respectively. In the presence of measurements, the domain wall will fluctuate away from its original path.}
\label{fig:fig2main}
\end{figure}

In terms of the statistical model, the quantum noises act as a magnetic field pining in the direction $\mathbb{I}$~\cite{BAO2021168618, Noise_bulk, jian2021quantum, PhysRevB.108.104310, PhysRevB.107.014307, PhysRevB.108.104310} and thus the weight of the spin configuration above for $F^{(n,k)}_{AB}$ with bulk spins in $\mathbb{C}$ is proportional to $d^{-qLT \vert \mathbb{C} \vert}$ and it is not favored anymore. Instead, quantum noises can relax the unitary constraints and induce other possible spin configurations. 
We leave the randomness of the locations of noises as a quenched disorder and show how to find the dominant spin configuration for each given trajectory of temporally uncorrelated quantum noises. As indicated in Fig. \ref{fig:fig2main} (b), spins remain $\mathbb{C}$ until the reversed evolution encounters a quantum noise $N(x_{1},t_{1})$. Spins inside the downward light cone of this quantum noise will change from $\mathbb{C}$ to $\mathbb{I}$ while other spins are unchanged. 
Other quantum noises inside the light cone of $N(x_{1}, t_{1})$ do not affect the spin configuration because the spins are already in the $\mathbb{I}$ domain, while another quantum noise outside the light cone, e.g., $N(x_{2},t_{2})$, will also change the spins within its respective backward light cone from $\mathbb{C}$ to $\mathbb{I}$. Consequently, the domain wall separating regions $\mathbb{C}$ and $\mathbb{I}$ is formed by the boundary of light cones as shown in Fig. \ref{fig:fig2main} (b) and can be regarded as a combination of many small domain walls with an effective length scale $L_{\text{eff}} \sim q^{-1}$ determined by the average distance between adjacent quantum noises. 

In the presence of the projective measurements, we also leave the randomness of the space-time locations as a quenched disorder. The projective measurements can be treated as random Gaussian potential and cause the fluctuation of the domain wall away from its original respective path. The free energy can be obtained by KPZ theory with $L_{\text{eff}} \sim q^{-1}$ and is consistent with the volume law entropy for the hybrid circuit \cite{Note1}. Although there are quantum noises present below the original domain wall with an average height $q^{-1}$, the average height of the fluctuated domain wall is $q^{-2/3}$ given by the KPZ theory with $L_{\text{eff}} \sim q^{-1}$ and thus we can neglect the effects of these quantum noises on the fluctuation of the domain wall.
For the mutual information $I_{A:B}$, the bulk terms proportional to the subsystem size cancel out, but the boundary term from 
the free energy of the domain wall near the midpoint is crucial because the effective length scale for $S_{A(B)}$ has been changed as shown in Fig. \ref{fig:fig2main} (a), resulting in
\begin{eqnarray} \label{eq:MI_main}
    I_{A:B}(q) \sim q^{-1/3}.
\end{eqnarray}
The theoretical prediction can be straightforwardly extended to temporally correlated quantum noise: it can be treated as emergent new boundaries that directly induce an effective length scale $L_{\text{eff}} \sim q^{-1}$~\cite{Note1}.

Furthermore, we consider the abilities of information protection in the steady state in the presence of quantum noises. 
One qudit information is encoded into the steady state by forming a Bell pair between a reference qudit and a middle qudit at $(x_{0}, t_{0})$ of the system, see Fig. \ref{fig:fig1main}. 
The encoded information can be measured by the mutual information between the system $AB$ and the reference qudit $R$ 
\begin{eqnarray}
    I_{AB:R}(t,q) = S_{AB}(t,q)+S_{R}(t,q)-S_{AB \cup R}(t,q).
\end{eqnarray}
Because the reference qudit and the middle qudit form a Bell pair, in the corresponding statistical model, the spin at position ($x_{0}, t_{0}$) is determined by the top boundary condition of the reference qudit. 
We also use $R$ to represent this Bell pair, which is fixed to $\mathbb{I}$, $\mathbb{C}$ and $\mathbb{C}$ for $S_{AB}$, $S_{R}$ and $S_{AB \cup R}$ respectively. 

Although the scalings of entanglement for temporally uncorrelated and correlated quantum noises are the same, the timescales of information protection are different. The dominant spin configuration for $F^{(n,k)}_{R}$ is where all the spins are $\mathbb{I}$ except the spin $R$ fixed to $\mathbb{C}$, therefore, $S_{R}$ is constant with contribution from the bubble created by $R_{\mathbb{C}}$. For the temporally correlated quantum noises, the dominant spin configurations for $F^{(n,k)}_{AB}$ and $F^{(n,k)}_{AB \cup R}$ change when the spin $R$ crosses the domain wall~\cite{Note1}. Therefore, the timescale is given by the average height of the domain wall which is $q^{-1}$ and $q^{-2/3}$ without and with monitored measurements.

On the contrary, for the temporally uncorrelated quantum noises, $R$ can act similarly to quantum noise and change the spins inside its light cone from $\mathbb{C}$ to $\mathbb{I}$ as shown in Fig. \ref{fig:fig2main} (c) and (d). 
Therefore, the domain wall configuration has been modified and the timescale of information protection does not correspond to the height of the domain wall.
A detailed analysis based on the statistical model is given in SM~\cite{Note1}.
This information protection process can also be understood as a Hayden-Preskill protocol~\cite{PatrickHayden_2007} as shown in Fig. \ref{fig:fig1main} (c). 
The steady state can be regarded as a black hole formed long ago that is maximally entangled with the environment, which is under the control of Bob. 
To destroy her recorded one-qudit information, which is maximally entangled with a reference qudit of Charlie, Alice throws it into the black hole. 
The quantum noise channels can be regarded as Hawking radiation to the environment. The Hayden-Preskill protocol tells us that the environment, i.e., Bob, only needs slightly more than one qudit from the Hawking radiation to decode Alice's information. Therefore, the timescale of information protection corresponds to the time required for a quantum noise with probability $q$ to appear in the light cone of the encoded information with area $O(t^{2})$, and hence the timescale is $q^{-1/2}$~\cite{Note1}. The presence of measurements will not alter the qualitative arguments above. Therefore, the timescales of information protection for temporally correlated and uncorrelated quantum noises in MIPT are $q^{-2/3}$ and $q^{-1/2}$, respectively. 
We can also apply statistical model understanding to the noiseless case, where the information can be protected by the subsystem of monitored circuits ($p_{m}<p_{m}^{c}$) forever or with timescale $L_{\rm{sub}}^{2/3}$ for $L_{\rm{sub}} > L/2$ or $L_{\rm{sub}} < L/2$ respectively~\cite{Note1}.

\textit{Clifford simulation.---} To support the theoretical predictions, we use QuantumClifford.jl package~\cite{quantumclifford} to perform extensive large-scale Clifford simulations~\cite{PhysRevA.70.052328, nielsen2002} where the random Clifford gates form a unitary 3-design~\cite{Unitary3design}. 
We use the reset channel $\mathcal{R}_{i}(\rho)=\tr_{i}(\rho) \otimes \vert 0 \rangle \langle 0 \vert_{i}$ to model the quantum noise,
which is easy to implement in the current generation of quantum hardware~\cite{reset_exp1, reset_exp2}, while our theoretical analysis does not depend on the choice of quantum channels. And we set the probability of projective measurement $0<p_{m}<p_{m}^{c}$. The numerical results with quantum dephasing channels and $p_{m}>p_{m}^{c}$ can be found in the SM~\cite{Note1}.

\begin{figure}[ht]
\centering
\includegraphics[width=0.49\textwidth, keepaspectratio]{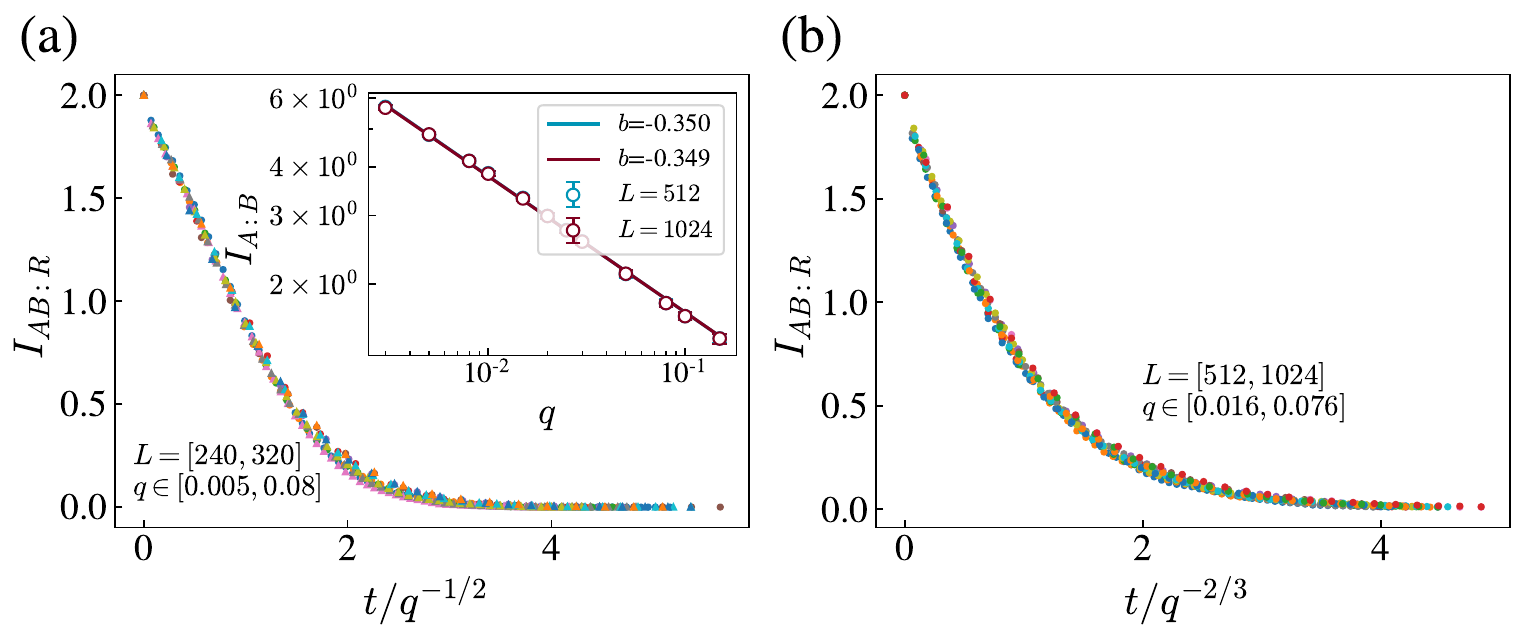}
\caption{The information protection dynamics. $q$ represents the probability of reset channels and $p_{m}=0.2<p_{m}^{c}$ is the probability of projective measurement. (a) shows the mutual information $I_{AB : R}$ vs rescaled time $t/q^{-1/2}$ for temporally uncorrelated quantum noises. The inset shows the fitting of mutual information $I_{A:B}$ with the function $I_{A:B}(q)=aq^{b}$. $b$ is very close to the theoretical prediction $-1/3$. (b) shows the mutual information $I_{AB : R}$ vs rescaled time $t/q^{-2/3}$ for temporally correlated quantum noises.}
\label{fig:fig3main}
\end{figure}

The dynamics of mutual information $I_{AB: R}$ in the presence of temporally uncorrelated and correlated quantum noises are shown in Fig. \ref{fig:fig3main} (see SM~\cite{Note1} for more numerical results for generic non-Markovian cases). 
The data with different system sizes and noise probabilities can be collapsed with rescaled time $t/q^{-1/2}$ and $t/q^{-2/3}$ for temporally uncorrelated and correlated quantum noises, consistent with the theoretical predictions. 
The fitting of the half-chain mutual information shown in the inset in Fig. \ref{fig:fig3main} (a) gives the $q^{-1/3}$ scaling. We have also demonstrated the scaling of von Neumann entropy~\cite{Note1}.

\textit{Discussions and conclusion.---} We provide a comprehensive analytical understanding of the impacts of quantum noises with different temporal correlations on entanglement generation and information protection: 
the mutual information satisfies the scaling $q^{-1/3}$ 
for both temporally uncorrelated and correlated noise, while the timescale of information protection for temporally uncorrelated and correlated noise is $q^{-1/2}$ and $q^{-2/3}$, respectively. 
These theoretical predictions are further demonstrated by convincing numerical results from Clifford circuit simulations~\cite{Note1}.

It is worth noting that the information protection capacity of non-Markovian noise (temporally correlated) is much stronger than the Markovian noise case when $q$ is small, which is consistent with recent studies on the unexpected benefits brought by non-Markovian noise in quantum simulation \cite{Chen2023a_z} and quantum computation \cite{Kattemolle2023_z}.
In the SM~\cite{Note1}, we also investigated the information protection capacity for a general non-Markovian noise interpolating between the Markovian and the strongest non-Markovian limits in the no measurement limit. 
We found that the information protection timescale changes continuously from $q^{-1/2}$ in the Markovian limit to $q^{-1}$ in the strong non-Markovian limit, consistent with the expectation.

Furthermore, our setup realizes the Hayden-Preskill protocol in hybrid quantum circuits by identifying the quantum noise channel (unitary evolution with ancilla qudits) as Hawking radiation through the ancilla qudits. The setup in this Letter overcomes the subtleties in the previous Hayden-Preskill analogy in hybrid quantum circuits \cite{Weinstein2022_z}. 
By coupling the reference qudit to the system after reaching the steady state, i.e. the black hole with half of the qudits radiated out and maximally entangled with the environment, we recover the Hayden-Preskill thought experiment, in which Bob can successfully decode the information, and hence substantially reduce the mutual information between the black hole and Charlie with $O(1)$ more qubits radiated.

In conclusion, we have presented a comprehensive theoretical framework for understanding entanglement generation and information protection in noisy hybrid quantum circuits. 
This framework is applicable to both temporally uncorrelated (Markovian) and correlated (non-Markovian) quantum noises. 
This work not only reveals a thorough theoretical understanding of $q^{-1/3}$ scaling in the presence of quantum noises on MIPT setup but also highlights the distinctions between the effects on information protection of quantum noises with different temporal correlations which are indistinguishable from the entanglement perspective and initial state information protection protocol~\cite{Coding_Vijay, liu2024noise, Note1}. Furthermore, our theoretical analysis can be extended to the cases of quantum noises with system size-dependent probability, where MIPT still exists and a new noise-induced entanglement phase transition has been investigated~\cite{liu2024noise}.

\textit{Acknowledgement.---} This work is supported in part by the NSFC under Grant No. 11825404 (SL, MRL).
SL and MRL acknowledge the support from the Lavin-Bernick Grant during their visit to Tulane University, where part of the work was conducted. The work of SKJ is supported by a startup fund at Tulane University. SXZ would like to acknowledge the helpful discussion with Yu-Qin Chen in terms of the non-Markovian noise.

%

\clearpage
\newpage
\widetext

\begin{center}
\textbf{\large Supplemental Material for ``Entanglement structure and information protection in noisy hybrid quantum circuits''}
\end{center}

\renewcommand{\thefigure}{S\arabic{figure}}
\setcounter{figure}{0}
\renewcommand{\theequation}{S\arabic{equation}}
\setcounter{equation}{0}
\renewcommand{\thesection}{\Roman{section}}
\setcounter{section}{0}
\setcounter{secnumdepth}{4}

\addtocontents{toc}{\protect\setcounter{tocdepth}{0}}
{
\tableofcontents
}

\section{Introduction to the effective statistical model}
In this section, we introduce the mapping framework between the hybrid quantum circuit and the effective classical statistical model and demonstrate how to obtain von Neumann entropy and mutual information of the circuit model from the free energy of the statistical model. Here, we begin with the most basic setup without quantum noises or projective measurements for the sake of simplicity and defer the discussions of cases involving quantum noises or projective measurements to the subsequent sections.

\subsection{The mapping between the quantum circuit and the statistical model}

\begin{figure}[ht]
\centering
\includegraphics[width=0.9\textwidth, keepaspectratio]{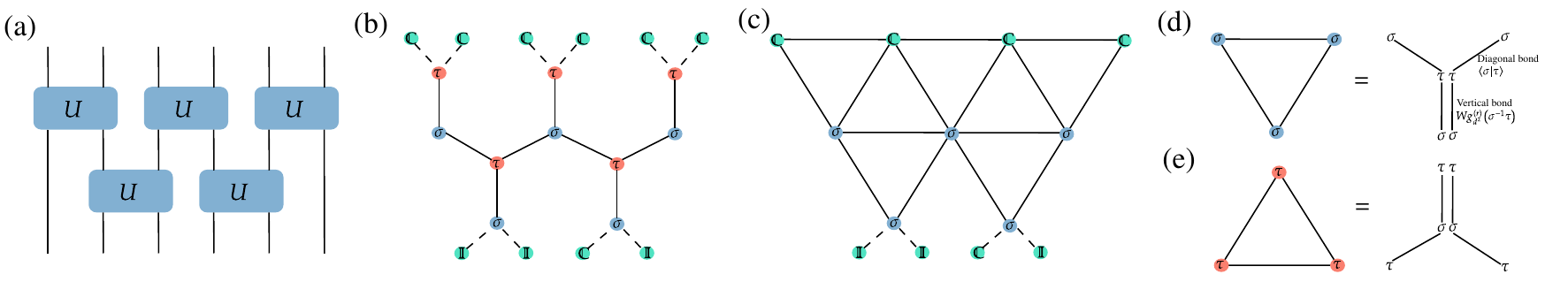}
\caption{(a) shows the quantum circuit with random two-qudit unitary gates arranged in a brick-wall structure. (b) shows the statistical model with the degrees of freedom formed by the permutation-valued spins $\sigma$ and $\tau$. The particular boundary conditions in correspondence with $F^{(n,k)}_{AB \cup R}$ is to add a layer of spins $\mathbb{C}$ at the top and to add a layer of spins $\mathbb{I}$ (maximally mixed state as the initial state) at the bottom except the spin $\mathbb{C}$ at position $x_{0}$. (c) we integrate out spins $\tau$ to obtain positive three-body weights of the downward triangles. (d) the weight of the triangle is the product of the weights of the diagonal bonds and the weight of the vertical bond. (e) when we choose the maximally mixed state as the initial state, we can also integrate out the spins $\sigma$ to obtain positive three-body weights of the upward triangles.}
\label{fig:Mapping}
\end{figure}

The quantum circuit is composed of random two-qudit unitary gates arranged in a brick-wall layered structure as shown in the main text. At discrete time step $T$, the density matrix $\rho$ is 
\begin{eqnarray}
    \rho = \prod_{t=1}^{T} \tilde{U}_{t} \rho_{0} \tilde{U}_{t}^{\dagger},
\end{eqnarray}
where $\rho_{0}$ is the density matrix of the initial state, and
\begin{eqnarray}
    \tilde{U}_{t}= \prod_{i=0}^{\frac{L-4}{2}} U_{t, (2i+2, 2i+3)}  \prod_{i=0}^{\frac{L-2}{2}} U_{t, (2i+1, 2i+2)}, 
\end{eqnarray}
is the unitary evolution of discrete time step $t$ where each two-qudit unitary gate is independently and randomly drawn from the Harr measure. To obtain the von Neumann entropy of the quantum circuit from the free energy of the statistical model as discussed below, we can express the density matrix in a $r$-fold replicated Hilbert space
\begin{eqnarray}
    \vert \rho \rangle^{\otimes r} &=& \prod_{t=1}^{T} \left[ \tilde{U}_{t} \otimes \tilde{U}_{t}^{*}  \right]^{\otimes r} \vert \rho_{0} \rangle^{\otimes r} \\ \nonumber 
    &=& \prod_{t=1}^{T} \left[ \prod_{i=0}^{\frac{L-4}{2}} (U_{t, (2i+2, 2i+3)} \otimes U^{*}_{t, (2i+2, 2i+3)})^{\otimes r}  \prod_{i=0}^{\frac{L-2}{2}} (U_{t, (2i+1, 2i+2)} \otimes U^{*}_{t, (2i+1, 2i+2)})^{\otimes r}  \right] \vert \rho_{0} \rangle^{\otimes r},
\end{eqnarray}
and the mapping to the effective statistical model arises from the average over the Haar random two-qudit unitary gates $U_{t,(i,j)}$~\cite{PhysRevB.101.104301, PhysRevB.101.104302, PhysRevX.7.031016, PhysRevLett.129.080501, PhysRevB.99.174205, PhysRevX.12.041002, collins2003moments, collinsIntegrationRespectHaar2006, PhysRevX.8.021014}:
\begin{eqnarray}
    \mathbb{E}_{\mathcal{U}}(U_{t,(i,j)} \otimes U_{t,(i,j)}^{*})^{\otimes r} = \sum_{\sigma, \tau \in S_{r}} \text{Wg}_{d^2}^{(r)}(\sigma \tau^{-1}) \vert \tau \tau \rangle \langle \sigma \sigma \vert_{ij},
\end{eqnarray}
where $S_{r}$ is the permutation group of dimension $r$ , $d$ is the local Hilbert space dimension of qudit ($d=2$ for qubit), and $\text{Wg}_{d^{2}}^{(r)}$ is the Weingarten function with an asymptotic expansion for large $d$~\cite{PhysRevB.99.174205, collinsIntegrationRespectHaar2006}:
\begin{eqnarray}
    \text{Wg}_{d^{2}}^{(r)} (\sigma)  = \frac{1}{d^{2r}} \left[ \frac{\text{Moeb}(\sigma)}{d^{2\vert\sigma\vert}} + \mathcal{O}(d^{-2\vert \sigma \vert -4})\right],
\end{eqnarray}
where $\vert \sigma \vert$ is the number of transpositions required to construct $\sigma$ from the identity permutation spin $\mathbb{I}$. Therefore, the quantum circuit has been transformed into a classical statistical model, where the degrees of freedom are formed by permutation-valued spins $\sigma$, $\tau$, see Fig. \ref{fig:Mapping} (a) and (b). 

The partition function $Z$ of this statistical model is obtained by summing the total weights of various spin configurations. For a specific spin configuration, the total weight is the product of the weights of the diagonal and vertical bonds where the weight of the diagonal bond is given by the inner product between two diagonally adjacent permutation spins
\begin{eqnarray}
    w_{d}(\sigma, \tau) = \langle \sigma \vert \tau \rangle = d^{r-\vert \sigma^{-1} \tau \vert},
\end{eqnarray}
and the weight of the vertical bond is determined by the Weingarten function. However, the $\text{Moeb}(\sigma)$ which is the Moebius number of $\sigma$ can be negative~\cite{collinsIntegrationRespectHaar2006}. To obtain positive definite weights, we can integrate out the $\tau$ spins and get positive three-body weights of downward triangles
\begin{eqnarray}
    W^{0}(\sigma_{1}, \sigma_{2}; \sigma_{3}) = \sum_{\tau \in S_{r}} \text{Wg}_{d^{2}}^{(r)} (\sigma_{3}\tau^{-1}) d^{2r-\vert \sigma_{1}^{-1} \tau \vert - \vert \sigma_{2}^{-1} \tau \vert},
    \label{eq:3body}
\end{eqnarray}
see Fig. \ref{fig:Mapping} (d) and (e) for details. Then the total weight of a specific spin configuration is the product of the weights of the downward triangles.
It is worth noting that we can also obtain the positive three-body weights of upward triangles by integrating out the $\sigma$ spins when the initial state is maximally mixed (see Fig. \ref{fig:Mapping} (e)). In the following discussion, we focus on the most dominant spin configuration that has the largest total weight in the large $d$ limit, i.e., the partition function $Z$ is determined by the weight of the dominant spin configuration.

Furthermore, we show that the spin-spin interaction in the statistical model is ferromagnetic. We can consider the weights of the downward triangles with some specific spin configurations.
\begin{itemize}
    \item $\sigma_{1} = \sigma_{2} = \sigma_{3} = \sigma$: 
    \begin{eqnarray}
    W^{0}(\sigma,\sigma; \sigma) &=& \sum_{\tau \in S_{r}} \text{Wg}_{d^{2}}^{(r)}(\sigma \tau^{-1}) d^{2r-2\vert \sigma^{-1} \tau \vert} \\ \nonumber
    &\approx& \sum_{\tau \in S_{r}} \text{Moeb}(\sigma \tau^{-1}) d^{-4 \vert \sigma^{-1} \tau \vert} \\ \nonumber
    &\approx& d^{0}.
    \end{eqnarray}
    \item $\sigma_{1} = \sigma^{\prime}, \sigma_{2}=\sigma_{3}=\sigma$ or $\sigma_{2} = \sigma^{\prime}, \sigma_{1}=\sigma_{3}=\sigma$:
    \begin{eqnarray}
        \label{eq:domainwall}
        W^{0}(\sigma^{\prime},\sigma; \sigma) = W^{0}(\sigma,\sigma^{\prime}; \sigma)&=& \sum_{\tau \in S_{r}} \text{Wg}_{d^{2}}^{(r)}(\sigma \tau^{-1}) d^{2r-\vert \sigma^{-1} \tau \vert-\vert (\sigma^{\prime})^{-1} \tau \vert} \\ \nonumber
        &\approx& \sum_{\tau \in S_{r}} \text{Moeb}(\sigma \tau^{-1}) d^{-3 \vert \sigma^{-1} \tau \vert - \vert (\sigma^{\prime})^{-1} \tau \vert} \\ \nonumber
        &\approx& d^{-\vert (\sigma^{\prime})^{-1}) \sigma \vert}.
    \end{eqnarray}
\end{itemize}
While there are other possible spin configurations, the configuration that maximizes the triangle weight occurs when $\sigma_{1} = \sigma_{2} = \sigma_{3} = \sigma$. Therefore, the spin-spin interaction is ferromagnetic, and all the spins tend to be in the same direction to achieve the largest total weight. However, as discussed below, due to the particular boundary conditions and the presence of quantum noises, the $S_{r}$ rotational symmetry is broken~\cite{BAO2021168618, Noise_bulk, jian2021quantum, PhysRevB.108.104310, PhysRevB.107.014307, PhysRevB.108.104310}
and domain walls may appear with unit energy of $\log (W^{0}(\sigma^{\prime},\sigma; \sigma))$. It is worth noting that the weight is zero when $\sigma_{1}=\sigma_{2} \neq \sigma_{3}$ known as unitary constraint~\cite{Uaverage_Qi, PRXQuantum.4.010331}. Therefore, when the domain wall passes through a triangle, it can only pass diagonally with spin configurations shown in Eq. \eqref{eq:domainwall}. Consequently, the domain wall is unique in the absence of quantum noises and measurements.

\subsection{The relation between the von Neumann entropy and the free energy}
In the last section, we have introduced the mapping between the hybrid quantum circuit and the ferromagnetic statistical model. Here, we show how to obtain von Neumann entropy of the circuit model from the free energy of the statistical model. Firstly, we can rewrite the von Neumann entropy $S_{\alpha}$ as
\begin{eqnarray}
    S_{\alpha} = \underset{n \rightarrow 1}{\lim} S^{(n)}_{\alpha} = \underset{n \rightarrow 1}{\lim} \frac{1}{1-n} \mathbb{E}_{\mathcal{U}} \log\frac{\tr \rho_{\alpha}^{n}}{(\tr \rho)^{n}},
\end{eqnarray}
where $\rho_{\alpha}$ is the reduced density matrix of region $\alpha$ and $S^{(n)}_{\alpha}$ is the $n$-th order Renyi entropy. As the mapping shown above, we can also represent $S^{(n)}_{\alpha}$ in $n$-fold replicated Hilbert space
\begin{eqnarray}
    S_{\alpha}^{(n)} = \frac{1}{1-n} \mathbb{E}_{\mathcal{U}} \log \frac{\tr \rho_{\alpha}^{n}}{(\tr \rho)^{n}} = \frac{1}{1-n} \mathbb{E}_{\mathcal{U}} \log \frac{\Tr((C_{\alpha} \otimes I_{\bar{\alpha}}) \rho^{\otimes n})}{\Tr ( (I_{\alpha} \otimes I_{\bar{\alpha}}) \rho^{\otimes n})} = \frac{1}{1-n} \mathbb{E}_{\mathcal{U}} \log \frac{Z^{(n)}_{\alpha}}{Z^{(n)}_{0}},
\end{eqnarray}
where $C = \begin{pmatrix} 1 & 2 & ... & n  \\ 2 & 3 & ... & 1\end{pmatrix} $ and $I = \begin{pmatrix} 1 & 2 & ... & n  \\ 1 & 2 & ... & n\end{pmatrix} $ are the cyclic and identity permutations in $n$-fold replicated Hilbert space, respectively. With the help of the  replica trick~\cite{nishimori2001statistical, kardar2007statistical}, we can overcome the difficulty of the average outside the logarithmic function
\begin{eqnarray}
    \mathbb{E}_{\mathcal{U}} \log Z_{\alpha}^{(n)} &=& \underset{k \rightarrow 0}{\lim} \frac{1}{k} \log \{ \mathbb{E}_{\mathcal{U}} (Z_{\alpha}^{(n)})^{k} \} = \underset{k \rightarrow 0}{\lim} \frac{1}{k} \log Z_{\alpha}^{(n,k)},\\ \nonumber 
    \mathbb{E}_{\mathcal{U}} \log Z_{0}^{(n)} &=& \underset{k \rightarrow 0}{\lim} \frac{1}{k} \log \{ \mathbb{E}_{\mathcal{U}} (Z_{0}^{(n)})^{k} \} = \underset{k \rightarrow 0 }{\lim} \frac{1}{k} \log Z_{0}^{(n,k)},
\end{eqnarray}
where 
\begin{eqnarray}
    \label{eq:topboundary}
    Z_{\alpha}^{(n,k)} &=&\Tr \left\{ ( C_{\alpha} \otimes I_{\bar{\alpha}}) ^{\otimes k} \left[ \mathbb{E}_{\mathcal{U}} \rho^{\otimes nk} \right] \right\} = \Tr  \left\{ \mathbb{C}_{\alpha} \otimes \mathbb{I}_{\bar{\alpha}}  \left[ \mathbb{E}_{\mathcal{U}} \rho^{\otimes nk} \right] \right\}, \\ \nonumber
    Z_{0}^{(n,k)} &=&\Tr \left\{ ( I_{\alpha} \otimes I_{\bar{\alpha}}) ^{\otimes k} \left[ \mathbb{E}_{\mathcal{U}} \rho^{\otimes nk} \right] \right\} = \Tr \left\{ \mathbb{I}_{\alpha} \otimes \mathbb{I}_{\bar{\alpha}}  \left[ \mathbb{E}_{\mathcal{U}} \rho^{\otimes nk} \right] \right\},
\end{eqnarray}
with $\mathbb{C} = \begin{pmatrix} 1 & 2 & ... & n  \\ 2 & 3 & ... & 1\end{pmatrix}^{\otimes k} $ and $\mathbb{I} = \begin{pmatrix} 1 & 2 & ... & n  \\ 1 & 2 & ... & n\end{pmatrix}^{\otimes k} $ are permutations in the $r$-fold replicated Hilbert space with $r=nk$. Therefore,
\begin{eqnarray}
    S_{\alpha} = \underset{n \rightarrow 1}{\underset{k \rightarrow 0 }{\lim}} \frac{1}{k(1-n)} \log \left\{ \frac{Z_{\alpha}^{(n,k)}}{Z_{0}^{(n,k)}} \right\},
    \label{eq:smzz}
\end{eqnarray}
where $Z$ is the partition function for the classical spin model via the mapping, and it corresponds to the weight of the dominant spin configuration with the largest weight of the ferromagnetic spin model with particular top boundary conditions in the large $d$ limit:
$\mathbb{C}_{\alpha} \otimes \mathbb{I}_{\bar{\alpha}}$ for $Z_{\alpha}$ and $\mathbb{I}_{\alpha} \otimes \mathbb{I}_{\bar{\alpha}}$ for $Z_{0}$. Therefore, $S_{\alpha}$ can be represented as the free energy difference:
\begin{eqnarray}
    S_{\alpha}^{(n,k)} = \frac{1}{k(n-1)} \left[ F_{\alpha}^{(n,k)} - F_{0}^{(n,k)}\right].
\end{eqnarray}
We note that the free energy $F^{(n,k)}$ is proportional to the length of the domain wall with unit energy $k(n-1)$ as shown below, and thus $\frac{1}{k(n-1)} F^{(n,k)}$ is independent of the index $(n,k)$. Consequently, the limit to extract von Neumann entropy shown in Eq.~\ref{eq:smzz} can be safely taken. Moreover, the above discussion has assumed that the initial state is a product state with a free bottom boundary condition. In the case with a maximally mixed state as the initial state, the bottom boundary is fixed to $\mathbb{I}_{\alpha} \otimes \mathbb{I}_{\bar{\alpha}}$ as discussed below.

\section{The analytical understanding in the presence of temporally uncorrelated quantum noises}
In this section, we present the theoretical predictions for the scalings of von Neumann entropy, mutual information, and the timescale of the encoded information protection in the presence of temporally uncorrelated quantum noises. Specifically, we first focus on the case in the presence of quantum dephasing channels with probability $q$ and defer the discussions for other quantum channels to the later part of this section.

\subsection{Quantum dephasing channel}
Quantum dephasing channel is one common choice to model the
quantum noise. Under the quantum dephasing channel acting on qudit $i$, the density matrix is
\begin{eqnarray}
    \rho^{\prime} = \mathcal{D}_{i}(\rho) = \sum_{j=0}^{d-1} P_{j} \rho P_{j},
    \label{eq:dephasing}
\end{eqnarray}
where $P_{j}=\vert j \rangle \langle j \vert$. Furthermore, it can be realized with an ancilla qudit in the environment initialized to $\vert 0 \rangle$ and unitary operation $V$
\begin{eqnarray}
    V=\sum_{i=1}^{d-1} (\vert ii \rangle  \langle i0 \vert + h.c. ) + \vert 00 \rangle  \langle 00 \vert + \sum_{j \neq 0, i} \vert ij \rangle  \langle ij \vert,
\end{eqnarray}
which is the CNOT gate when $d=2$ with the ancilla qubit as the target qubit. The density matrix of these two qudits is
\begin{eqnarray}
    V \rho \otimes \vert 0 \rangle \langle 0 \vert V^{\dagger} &=&  \sum_{i=0}^{d-1} \vert ii \rangle  \langle i0 \vert  \left[ \rho \otimes \vert 0 \rangle \langle 0 \vert  \right] \sum_{i=0}^{d-1} \vert i0 \rangle  \langle ii \vert \\ \nonumber
    &=& \sum_{ij} \rho_{ij} \vert ii \rangle \langle jj \vert.
\end{eqnarray}
Therefore, if we discard (trace out) the ancilla qudit in the environment, the reduced density matrix is the same as that in Eq. \eqref{eq:dephasing}. This is the reason why the quantum channel can be regarded as Hawking radiation as discussed in the main text. Moreover, the analogy to the Hayden-Preskill protocol is suitable for other quantum channels because they can also be realized with the introduced ancilla qudit in the environment and unitary operation~\cite{nielsen2010quantum}.

In terms of the statistical model, the quantum dephasing channels modify the inner product between adjacent spins thereby affecting the three-body weights with the same spin configurations. The exact results of the inner product of $\langle \sigma \vert \mathcal{D} \vert \tau \rangle$ with general $\sigma$ and $\tau$ are hard to track~\cite{PhysRevLett.129.080501}. However, we know that
\begin{eqnarray}
    \langle \sigma \vert \mathcal{D} \vert \sigma \rangle = \langle \sigma \vert \mathcal{D} \vert \mathbb{I} \rangle = \langle \mathbb{I} \vert \mathcal{D} \vert \sigma \rangle = d^{r-\vert \sigma \vert },
\end{eqnarray}
and the inner products with other spin configurations are smaller by at least a factor of $1/d$. In the absence of quantum noises
\begin{eqnarray}
    W^{0}(\mathbb{C}, \mathbb{C}; \mathbb{C}) = W^{0}(\mathbb{I}, \mathbb{I}; \mathbb{I}) = d^{0}.
\end{eqnarray}
while if there is a quantum dephasing channel between $\sigma_{1}(\sigma_{2})$ and $\tau$, 
\begin{eqnarray}
    W^{\mathcal{D}}(\mathbb{C}, \mathbb{C}; \mathbb{C}) &=& \sum_{\tau \in S_{r}} \text{Wg}^{(r)}_{d^{2}} (\mathbb{C}^{-1}\tau) d^{r-\vert \mathbb{C}^{-1} \tau \vert } \langle \mathbb{C} \vert \mathcal{D} \vert \tau \rangle \sim d^{-\vert \mathbb{C} \vert},
    \label{eq:WD_C}
\end{eqnarray}
\begin{eqnarray}
    W^{\mathcal{D}}(\mathbb{I}, \mathbb{I}; \mathbb{I}) &=& \sum_{\tau \in S_{r}} \text{Wg}^{(r)}_{d^{2}} (\mathbb{I}^{-1}\tau) d^{r-\vert \mathbb{I}^{-1} \tau \vert } \langle \mathbb{I} \vert \mathcal{D} \vert \tau \rangle \sim d^{0}.
    \label{eq:WD_I}
\end{eqnarray}
Therefore, the quantum noise in the region $\mathbb{C}$ reduces the three-body weight and induces a local energy cost similar to the magnetic field pinning in the $\mathbb{I}$ direction~\cite{BAO2021168618, Noise_bulk, jian2021quantum, PhysRevB.108.104310, PhysRevB.107.014307, PhysRevB.108.104310}. Besides these two spin configurations, other possible spin configurations are crucial for entanglement generation and information protection as discussed below. 

\subsection{The timescale of the information protection: maximally mixed state}
In the presence of the temporally uncorrelated quantum dephasing channels, the steady state is the maximally mixed state $\rho = \frac{I}{d^{L}}$. To gain insights into the analytical understanding of the timescale of the information protection, we consider a setup where the initial state is the maximally mixed state and one-qudit information is encoded into the system ($AB$) via coupling a reference qudit ($R$) with the qudit at position $x_{0}$ and time $0$. We will provide the theoretical prediction of the timescale required for $I_{AB : R}$ decaying to zero. 

The von Neumann entropy is the free energy difference
\begin{eqnarray}
    S_{\alpha}^{(n,k)} = \frac{1}{k(n-1)} \left[ F_{\alpha}^{(n,k)} - F_{0}^{(n,k)}\right]
\end{eqnarray}
of the spin models with particular top boundary conditions as shown in Eq. \eqref{eq:topboundary}. For $F_{\alpha}^{(n,k)}$, the top boundary conditions correspond to add a layer of spins $\mathbb{C}$, $\mathbb{I}$, and $\mathbb{C}$ for $\alpha = AB, R, AB \cup R$, respectively. The initial maximally mixed state corresponds to adding a layer of spins $\mathbb{I}$ at the bottom, except the spin at position $x_{0}$ which is fixed to $\mathbb{I}$, $\mathbb{C}$, and $\mathbb{C}$ for $\alpha=AB, R, AB \cup R$ respectively determined by the top boundary conditions of the reference qudit. For $F_{0}^{(n,k)}$, the spins on the additional top and bottom layers are all fixed to $\mathbb{I}$. We show the particular boundary conditions for $F_{AB \cup R}^{(n,k)}$ in Fig. \ref{fig:Mapping} (c). It is worth noting that each $\sigma$ or $\tau$ spin corresponds to two sites, i.e., $\vert \sigma \sigma \rangle$ or $\langle \tau \tau \vert$, see Fig. \ref{fig:Mapping} (d) and (e) for more details. We leave the randomness of the locations of quantum noises as a quenched disorder.

The calculation of $F_{0}^{(n,k)}$ is simple. We integrate out the $\tau$ spins to obtain positive three-body weights of the downward triangles. The spins on the top layer can be combined with the $\sigma$ spins on the nearest layer to generate positive three-body weights. Due to the fixed spins $\mathbb{I}$ on the top layer and the quantum dephasing channels in the bulk pinning in the $\mathbb{I}$ direction, the dominant spin configuration is the one where all the bulk $\sigma$ spins are $\mathbb{I}$. The partition function is the product of the weights of triangles ($W^{0}(\mathbb{I}, \mathbb{I}; \mathbb{I})=d^{0}$) and the weights of the diagonal bonds at the bottom, $Z^{(n,k)}_{0} = \langle \mathbb I \vert \mathbb{I} \rangle^{L} = d^{rL}$. Therefore, the free energy is $F^{(n,k)}_{0}= -\log Z^{(n,k)}_{0} = -rL \log(d)$.

Now, we consider the calculation of $F^{(n,k)}_{\alpha}$.
\begin{itemize}
    \item $F^{(n,k)}_{AB}$: we integrate out the $\sigma$ spins to obtain positive three-body weights of the upward triangles. In this case, the spins on the bottom layer can be combined with the $\tau$ spins on the nearest layer to generate positive three-body weights. Due to the fixed spins $\mathbb{I}$ on the bottom layer and the existence of quantum dephasing channels pining in the $\mathbb{I}$ direction, the dominant spin configuration is the one where all the bulk $\tau$ spins are $\mathbb{I}$, see Fig. \ref{fig:mixedstate} (a). The partition function is the product of the weights of triangles and the weights of the diagonal bonds at the top, $Z^{(n,k)}_{AB}=\langle \mathbb{C} \vert \mathbb{I} \rangle^{L} = d^{(r-\vert \mathbb{C} \vert)L}$. Therefore, the free energy is $F^{(n,k)}_{AB} = -(r-\vert \mathbb{C} \vert)L \log(d)$ and $S_{AB}= L \log(d)$ because $\vert \mathbb{C} \vert = k(n-1)$.
    
    \item $F^{(n,k)}_{R}$: we keep downward triangles by tracing $\tau$ spins and the dominant spin configuration is the one where all the bulk $\sigma$ spins are $\mathbb{I}$. The partition function is the product of the weights of triangles and the weights of the diagonal bonds at the bottom, $Z^{(n,k)}_{R} = \langle \mathbb{I} \vert \mathbb{I} \rangle^{L-1} \langle \mathbb{C} \vert \mathbb{I} \rangle = d^{rL-\vert \mathbb{C} \vert}$. Therefore, $F^{(n,k)}_{R} = - (rL-\vert \mathbb{C} \vert) \log(d)$ and $S_{R}= \log(d)$.

    \item $F^{(n,k)}_{AB \cup R}$: we integrate out the $\sigma$ spins to obtain positive three-body weights of upward triangles. 
    \begin{itemize}
    \item Firstly, we consider the case without any quantum dephasing channels. The initially fixed spin $\mathbb{C}$ at position $x_{0}$ can be denoted as $R_{\mathbb{C}}(x_{0},0)$. 
    \begin{itemize}
        \item Scenario I: all bulk $\tau$ spins are $\mathbb{I}$, see Fig. \ref{fig:mixedstate} (c). Consequently, the total weight is $\langle \mathbb{C} \vert \mathbb{I} \rangle^{L} W^{0}(\mathbb{C},\mathbb{I};\mathbb{I}) = d^{(r-\vert \mathbb{C} \vert) L} d^{-\vert \mathbb{C} \vert}=d^{rL-\vert \mathbb{C} \vert (L+1)}$ due to the spins $\mathbb{C}$ on the top layer and $R_{\mathbb{C}}(x_{0},0)$ on the bottom layer.
        \item Scenario II: the spins inside the light cone of $R_{\mathbb{C}}(x_{0},0)$ are $\mathbb{C}$ and the other bulk spins are $\mathbb{I}$ (see Fig. \ref{fig:mixedstate} (d)). The total weight is $W^{0}(\mathbb{C}, \mathbb{I}; \mathbb{I})^{2T-1} \langle \mathbb{C} \vert \mathbb{C} \rangle^{2T} \langle \mathbb{C} \vert \mathbb{I} \rangle^{L-2T} = d^{-\vert \mathbb{C} \vert(2T-1)} d^{2Tr}d^{(r-\vert \mathbb{C} \vert)(L-2T)}=d^{rL-\vert \mathbb{C} \vert(L-1)}$, where $T$ is the evolution time. The power $2T-1$ is because the propagation distance of the unitary gate next to $R_{\mathbb{C}}(x_{0},0)$ is $2$, not $1$, see Fig. \ref{fig:Distance} for details.
        \begin{figure}[ht]
        \centering
        \includegraphics[width=0.45\textwidth, keepaspectratio]{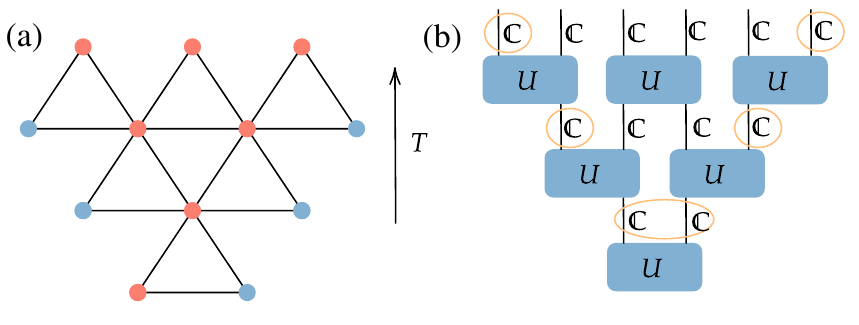}
        \caption{(a) the spin configuration of Scenario II. The left spin in the bottom triangle represents the spin $R$ fixed to $\mathbb{C}$. (b) shows the corresponding quantum circuit. Only $2T-1$ triangles contribute to the total weight because the propagation distance of the lowest triangle, i.e., the lowest unitary gate, is 2.}
        \label{fig:Distance}
        \end{figure}
    \end{itemize}
    The spin configuration in Scenario II is dominant and thus $S_{AB \cup R}=(L-1) \log (d)$ and $I_{AB : R} = 2 \log(d)=2\; (d=2)$. The encoded information is perfectly protected in the absence of quantum noises.

    \item  In the presence of quantum dephasing channels located outside the light cone of $R_{\mathbb{C}}(x_{0},0)$, the dominant spin configuration and the total weights remain unchanged, i.e., the information can still be protected. 
    
    \item  The free energy difference between the spin configurations in Scenarios I and II is $2 \vert \mathbb{C} \vert \log (d)$. Therefore, if more than two quantum dephasing channels are present within the light cone of $R_{\mathbb{C}}(x_{0},0)$ ($W^{\mathcal{D}}(\mathbb{C}, \mathbb{C}; \mathbb{C}) = d^{-\vert \mathbb{C} \vert}$), the dominant spin configuration will switch to that of Scenario I. Consequently, the mutual information $I_{AB : R} = 0.0$, i.e., the encoded information is destroyed.
    \end{itemize}

\end{itemize}

\begin{figure}[ht]
\centering
\includegraphics[width=0.8\textwidth, keepaspectratio]{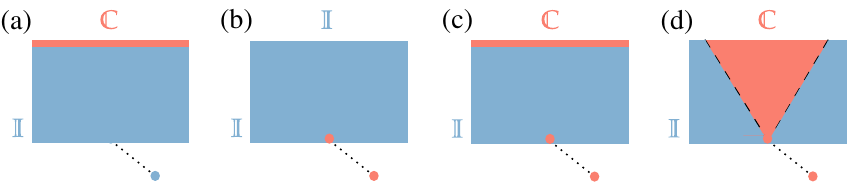}
\caption{(a) shows the spin configurations of $F^{(n,k)}_{AB}$ (b) shows the spin configuration of $F^{(n,k)}_{R}$, (c) shows the spin configuration of $F^{(n,k)}_{AB \cup R}$ in Scenario I, (d) shows the spin configuration of $F^{(n,k)}_{AB \cup R}$ in Scenario II. The dashed line in (d) represents the light cone of $R_{\mathbb{C}}(x_{0},0)$. }
\label{fig:mixedstate}
\end{figure}

Based on the theoretical analysis above, the mutual information $I_{AB : R}(t, q)$ at time $t$ and with quantum dephasing channel probability $q$ averaged over random space-time locations is:
\begin{eqnarray}
    I_{AB : R}(t,q) &=& 2.0 * (1-q)^{t(t+1)} + 1.0 * C_{t(t+1)}^{1} q(1-q)^{t(t+1)-1} \\ \nonumber
    &=& 2(1-q)^{t(t+1)} + t(t+1) q(1-q)^{t(t+1)-1},
\end{eqnarray}
where $t(t+1) = \sum_{i=1}^{t}2i$ is the area of the light cone. To determine the timescale of information protection, the time required for $I_{AB : R}$ decay to zero, we can first consider the limit as $q \rightarrow 0$
\begin{eqnarray}
    I_{AB : R}(t,q) \approx 2 + (-t-t^{2}) q+ \mathcal{O}(q^{3}),
\end{eqnarray}
indicating that the timescale of information protection is
\begin{eqnarray}
    t \approx \sqrt{2}q^{-1/2}-\frac{1}{2}+\frac{\sqrt{q}}{8\sqrt{2}} + \mathcal{O}(q^{3/2}).
\end{eqnarray}
Namely, the encoded information is destroyed after the time scale $q^{-1/2}$.
Therefore, we can substitute the time $t$ with $t_{0}q^{-1/2}$, 
\begin{eqnarray}
    I_{AB : R}(t_{0}q^{-1/2}, q) &=& 2(1-q)^{t_{0}^{2}q^{-1} +  t_{0}q^{-
1/2}}  + \left[ t_{0}^{2} + t_{0}q^{1/2}  \right](1-q)^{t_{0}^2q^{-1}+t_{0}q^{-1/2}-1}\\ \nonumber
    &=& 2 \left[ (1-q)^{q^{-1}} \right]^{t_{0}^{2}}  \left[ (1-q)^{q^{-1/2}} \right]^{t_{0}} +  \left[ t_{0}^{2} + t_{0}q^{1/2}  \right] \left[ (1-q)^{q^{-1}}\right]^{t_{0}^{2}} \left[ (1-q)^{q^{-1/2}}\right]^{t_{0}}(1-q)^{-1} \\ \nonumber
    & \underset{q \rightarrow 0}{\approx} & 2e^{-t_{0}^{2}} + t_{0}^{2} e^{-t_{0}^{2}},
\end{eqnarray}
which is independent of the quantum noise probability $q$. Consequently, the dynamics of $I_{AB : R}$ should be collapsed with rescaled time $t_{0} = t/q^{-1/2}$. 

Furthermore, this information protection process can also be understood as a Hayden-Preskill protocol~\cite{PatrickHayden_2007} as discussed in the main text which gives the same $q^{-1/2}$ scaling for the timescale of information protection.

\subsection{The timescale of the information protection: product state as the initial state}
Now, we consider the setup where the initial state is chosen as a product state as discussed in the main text. In this setup, a reference qudit is maximally entangled with the qudit at position $x_{0}$ after the system reaches the steady state to encode one-qudit quantum information. Then, we detect the dynamics of the mutual information $I_{AB : R}$. In the absence of the projective measurements, this setup is physically equivalent to the setup discussed above. However, due to the free bottom boundary determined by the initial product state, 
\begin{eqnarray}
    S^{n,k}_{\alpha} = \frac{1}{k(n-1)} (F^{(n,k)}_{\alpha} - F^{(n,k)}_{0}) =  \frac{1}{k(n-1)} F^{(n,k)}_{\alpha},
\end{eqnarray}
and the analysis in terms of the statistical model is slightly different.  Moreover, we must integrate out $\tau$ spins to obtain the positive three-body weights.

\begin{figure}[ht]
\centering
\includegraphics[width=0.48\textwidth, keepaspectratio]{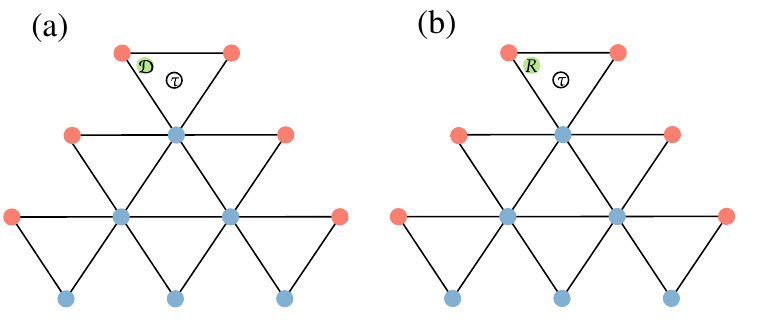}
\caption{(a), there is a quantum dephasing channel between $\sigma_{1}$ and $\tau$ in the top triangle; (b), there is the spin forming the Bell pair between $\sigma_{1}$ and $\tau$ in the top triangle. This spin is fixed to $\mathbb{C}$ or $\mathbb{I}$ given by the top boundary condition of the reference qubit.}
\label{fig:dephasing_and_bell}
\end{figure}

Due to the presence of quantum noises in bulk,  which act like magnetic field pinning in the direction $\mathbb{I}$, and the fixed top boundary spins $\mathbb{C}$ in the region $\alpha$, a domain wall must be formed to separate the regions $\mathbb{C}$ and $\mathbb{I}$. To gain insights into when and where the domain wall is created, we can estimate the three-body weights of downward triangles in the presence of a quantum dephasing channel (Fig. \ref{fig:dephasing_and_bell} (a)) or the spin forming the Bell pair (Fig. \ref{fig:dephasing_and_bell} (b)) with specific spin configurations before the discussion on the free energy. 

\begin{itemize}
    \item Dephasing channel: in the absence of quantum dephasing channels, the spin configuration shown in Fig. \ref{fig:dephasing_and_bell} (a) where the red circle represents spin $\mathbb{C}$ and blue circle represents spin $\mathbb{I}$) is forbidden by the unitary constraint $W^{0}(\mathbb{C}, \mathbb{C}; \mathbb{I})=0$~\cite{Uaverage_Qi, PRXQuantum.4.010331}. However, the presence of a quantum dephasing channel can break this unitary constraint. In the large $d$ limit, the three-body weight $W^{\mathcal{D}}(\mathbb{C}, \mathbb{C}; \mathbb{I})$ is:
    \begin{eqnarray}
        W^{\mathcal{D}}(\mathbb{C}, \mathbb{C}; \mathbb{I}) &=& \sum_{\tau \in S_{r}} \text{Wg}^{(r)}_{d^{2}} (\mathbb{I}^{-1}\tau) d^{r-\vert \mathbb{C}^{-1} \tau \vert } \langle \mathbb{C} \vert \mathcal{D} \vert \tau \rangle \\ \nonumber 
        & \sim & d^{-2\vert \mathbb{C} \vert}, 
    \end{eqnarray}
    with $\tau =  \mathbb{I}$. To verify this, we calculate the weight with $r=2$: $W^{\mathcal{D}}(\mathbb{C}, \mathbb{C}; \mathbb{I})=  \frac{d^{2}-d}{d^{4}-1} \sim d^{-2\vert \mathbb{C} \vert}$. Compared with the weight $ W^{\mathcal{D}}(\mathbb{C}, \mathbb{C}; \mathbb{C}) \sim d^{-\vert \mathbb{C} \vert}$ shown in Eq. \eqref{eq:WD_C},  although the spin configuration of $(\mathbb{C}, \mathbb{C}; \mathbb{I})$ can exist, its weight is less than that of $(\mathbb{C}, \mathbb{C}; \mathbb{C})$. It seems like the spin configuration of $(\mathbb{C}, \mathbb{C}; \mathbb{C})$ is dominant and we can treat the quantum noises as a local magnetic field as discussed in the last section and in previous works. However, the non-local effects of quantum noises have been neglected before, where the quantum dephasing channels near the top boundary can alter the spin configuration from $\mathbb{C}$ to $\mathbb{I}$ to exclude other quantum dephasing channels outside of the region $\mathbb{C}$ to minimize the free energy cost. Consequently, the spin configuration shown in Fig. \ref{fig:dephasing_and_bell} (a), where the spins inside the downward light cone of the quantum dephasing channel fixed to $\mathbb{I}$ is favored, despite having a lower local weight. See more details below.

    \item Bell pair: the spin forming the Bell pair denoted as $R$ is fixed to $\mathbb{I}$ for $F^{(n,k)}_{AB}$ and $\mathbb{C}$ for $F^{(n,k)}_{AB \cup R}$ and $F^{(n, k)}_R$, respectively. 
    \begin{itemize}
        \item The spin $R$ is $\mathbb{C}$ ($R_{\mathbb{C}}$) due to boundary condition. 
        \begin{itemize}
            \item $R_{\mathbb{C}}$ resides in region $\mathbb{C}$,
        \begin{eqnarray}
            W^{R_{\mathbb{C}}}(\mathbb{C}, \mathbb{C}; \sigma) = \sum_{\tau \in S_{r}} \text{Wg}^{(r)}_{d^{2}} (\sigma^{-1}\tau)  d^{r-\vert \mathbb{C}^{-1} \tau \vert} \langle \mathbb{C} \vert \mathbb{C} \rangle,  
        \end{eqnarray}
        with $W^{R_{\mathbb{C}}}(\mathbb{C}, \mathbb{C}; \mathbb{C}) \sim d^{0}$ and $W^{R_{\mathbb{C}}}(\mathbb{C}, \mathbb{C}; \mathbb{I}) \sim d^{-\vert \mathbb{C} \vert}$.
          \item $R_{\mathbb{C}}$ resides in region $\mathbb{I}$,
        \begin{eqnarray}
            W^{R_{\mathbb{C}}}(\mathbb{I}, \mathbb{I}; \mathbb{I}) = \sum_{\tau \in S_{r}} \text{Wg}^{(r)}_{d^{2}} (\mathbb{I}^{-1}\tau)  d^{r-\vert \mathbb{I}^{-1} \tau \vert} \langle \mathbb{I} \vert \mathbb{C} \rangle \sim d^{-\vert \mathbb{C} \vert}.  
        \end{eqnarray}
        \end{itemize}

        \item The spin $R$ is $\mathbb{I}$ ($R_{\mathbb{I}}$) due to boundary condition. 
        \begin{itemize}
            \item $R_{\mathbb{I}}$ resides in region $\mathbb{C}$,
        \begin{eqnarray}
            W^{R_{\mathbb{I}}}(\mathbb{C}, \mathbb{C}; \sigma) = \sum_{\tau \in S_{r}} \text{Wg}^{(r)}_{d^{2}} (\sigma^{-1}\tau)  d^{r-\vert \mathbb{C}^{-1} \tau \vert} \langle \mathbb{C} \vert \mathbb{I} \rangle,
        \end{eqnarray}
        with $W^{R_{\mathbb{I}}}(\mathbb{C}, \mathbb{C}; \mathbb{C}) \sim d^{-\vert \mathbb{C} \vert}$ and $W^{R_{\mathbb{I}}}(\mathbb{C}, \mathbb{C}; \mathbb{I}) \sim d^{-2\vert \mathbb{C} \vert}$. 
            \item $R_{\mathbb{I}}$ resides in region $\mathbb{I}$,
        \begin{eqnarray}
            W^{R_{\mathbb{I}}}(\mathbb{I}, \mathbb{I}; \mathbb{I}) = \sum_{\tau \in S_{r}} \text{Wg}^{(r)}_{d^{2}} (\mathbb{I}^{-1}\tau)  d^{r-\vert \mathbb{I}^{-1} \tau \vert} \langle \mathbb{I} \vert \mathbb{I} \rangle \sim d^{0}.  
        \end{eqnarray}
        \end{itemize}
        Similar to the case of the quantum dephasing channel, the spin configuration with the spins inside the respective light cone of spin $R$ fixed to $\mathbb{I}$ as illustrated in Fig, \ref{fig:dephasing_and_bell} (b) is favored although its weight is smaller locally. 
        \end{itemize}
\end{itemize}

Based on the analysis of the weights of triangles with specific spin configurations, we now present the theoretical analysis of the free energy with a product state as the initial state.

\begin{figure}[ht]
\centering
\includegraphics[width=0.6\textwidth, keepaspectratio]{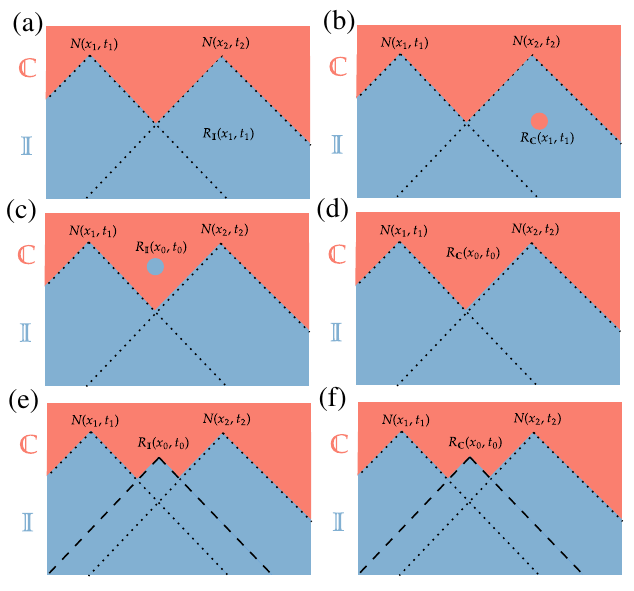}
\caption{The spin configurations of $F^{(n,k)}_{AB}$ and $F^{n,k}_{AB \cup R}$ with a product state as the initial state. (a)(b), late-time spin configurations, i.e., the spin $R$ forming the Bell pair is deep. (c)(d), early-time spin configurations in Scenario I, i.e., the spin $R$ forming the Bell pair is shallow. (e)(f), early-time spin configurations in Scenario II. Under the assumption that there is no quantum dephasing channel inside the upward light cone of the spin $R$, the spin configurations shown in (e)(f) are dominant compared with those shown in (c)(d). In the presence of quantum dephasing channels inside the upward light cone of the spin $R$, the spin configurations shown in (a)(b) are dominant, which is similar to the case with the spin $R$ is deep.}
\label{fig:productstate}
\end{figure}

\begin{itemize}
    \item $F^{(n,k)}_{R}$: the dominant spin configuration is that all the spins are $\mathbb{I}$. Due to the fixed spin $R_{\mathbb{C}}$ at position $x_{0}$ and time $t_{0}$, denoted as $R_{\mathbb{C}}(x_{0}, t_{0})$, the partition function is $Z^{(n,k)}_{R} = W^{R_{\mathbb{C}}}(\mathbb{I}, \mathbb{I}; \mathbb{I})= d^{-\vert \mathbb{C} \vert}$. Therefore, $F^{(n,k)}_{R} = \vert \mathbb{C} \vert \log(d)$ and thus $S_{R}=\log (d)$.

    \item $F^{(n,k)}_{AB}$ and $F^{(n,k)}_{AB \cup R}$: we consider the reversed evolution from the top to the bottom to analyze the formation of the domain wall in their respective dominant spin configurations.
    \begin{itemize}
        \item In the absence of the Bell pair, i.e., the spin $R$ fixed to $\mathbb{I}$ $(R_{\mathbb{I}}(x_{0},t_{0}))$ for $F_{AB}$ or $\mathbb{C}$ $(R_{\mathbb{C}}(x_{0},t_{0}))$ for $F_{AB \cup R}$. Due to the fixed spins $\mathbb{C}$ at the top boundary, the spins in the bulk will remain in the direction $\mathbb{C}$ until the evolution encounters a quantum dephasing channel at position $x_{1}$ and time $t_{1}$ denoted as $N(x_{1},t_{1})$. In the case where there is only this one quantum dephasing channel, the boundary of the light cone of $N(x_{1},t_{1})$, denoted as $c(x_{1}, t_{1})$, will form the domain wall separating the regions $\mathbb{C}$ and $\mathbb{I}$. If there are other quantum dephasing channels inside $c(x_{1},t_{1})$, the spins surrounding these new quantum dephasing channels are already $\mathbb{I}$, therefore, they have no effects. In the presence of a quantum dephasing channel at position $x_{2}$ and time $t_{2}$ outside $c(x_{1}, t_{1})$, the spins inside the light cone of $N(x_{2}, t_{2})$, denoted as $c(x_{2}, t_{2})$, are altered to $\mathbb{I}$. Consequently, the domain wall is formed by the boundary of light cone $c_{x_{1}, t_{1}} \cup c_{x_{2}, t_{2}}$. Therefore, in general cases, the domain wall is the boundary of the light cone $c_{x_{1}, t_{1}} \cup c_{x_{2}, t_{2}} \cup ... \cup c_{x_{N_{d}}, t_{N_{d}}}$, where $N_{d}$ is the number of quantum dephasing channels along the domain wall. These quantum dephasing channels are not within the light cones of each other. The total weight given by the domain wall is 
        \begin{eqnarray}
            W_{DW} = W^{0}(\mathbb{C}, \mathbb{I}; \mathbb{I})^{L-2N_{d}}  W^{\mathcal{D}}(\mathbb{C}, \mathbb{C}; \mathbb{I})^{N_{d}} = d^{-\vert \mathbb{C} \vert (L-2N_{d})} d^{-2\vert \mathbb{C} \vert N_{d}} = d^{-\vert \mathbb{C} \vert L}, 
        \end{eqnarray}
        where $W^{\mathcal{D}}(\mathbb{C}, \mathbb{C}; \mathbb{I})^{N_{d}}$ is the product of weights of the triangles including a quantum dephasing channel (the top triangle shown in Fig. \ref{fig:dephasing_and_bell} (a)) and $W(\mathbb{C}, \mathbb{I}; \mathbb{I})^{L-2N_{d}}$  is the total weight of the domain wall excluding the contribution from the triangles including a quantum dephasing channel whose propagation distance is $2$. It is worth noting that this total weight does not depend on $N_{d}$. As mentioned above, $W^{\mathcal{D}}(\mathbb{C}, \mathbb{C}; \mathbb{I}) < W^{\mathcal{D}}(\mathbb{C}, \mathbb{C}; \mathbb{C})$. A natural question arises as to why the spin configuration changes from $\mathbb{C}$ to $\mathbb{I}$ immediately upon encountering a quantum dephasing channel, rather than remaining in the direction $\mathbb{C}$ and incurring a local magnetic field energy cost. Due to the particular boundary conditions and the existence of extensive quantum dephasing channels, there must be a domain wall separating regions $\mathbb{C}$ and $\mathbb{I}$ with total weight $W_{DW}$. If there are other quantum dephasing channels residing in region $\mathbb{C}$, the total weight is  $W_{DW} W^{\mathcal{D}}(\mathbb{C}, \mathbb{C}; \mathbb{C})^{N_{d}^{\prime}} <  W_{DW}$, where $N_{d}^{\prime}$ is the number of dephasing channels residing in region $\mathbb{C}$. Therefore, the dominant spin configuration is to form a domain wall that is induced by the quantum dephasing channels near the top boundary such that all other quantum dephasing channels reside in the region $\mathbb{I}$ with no energy cost.
        
        \item The spin $R$ is deep (see Fig. \ref{fig:productstate}) (a) and (b)), i.e., far away from the top boundary. In this case, the spin $R$ must reside in the region $c_{x_{1}, t_{1}} \cup c_{x_{2}, t_{2}} \cup ... \cup c_{x_{N_{d}}, t_{N_{d}}}$. Therefore, the partition functions are the products of the weight of the domain wall and the weight of the triangle including the spin $R$, $Z^{(n,k)}_{AB}=W_{DW}W^{R_{\mathbb{I}}}(\mathbb{I}, \mathbb{I}; \mathbb{I}) = d^{-\vert \mathbb{C} \vert L}$ and $Z^{(n,k)}_{AB \cup R} = W_{DW} W^{R_{\mathbb{C}}}(\mathbb{I}, \mathbb{I}; \mathbb{I} ) = d^{-\vert \mathbb{C} \vert (L+1)}$. Therefore, the mutual information is $I_{AB : R} = 0$, i.e., the encoded information has been destroyed.
        \item The spin $R$ is shallow (see Fig. \ref{fig:productstate}) (c) and (d)). There are two scenarios:
        \begin{itemize}
            \item Scenario I: the spin $R$ resides in the region $\mathbb{C}$ as shown in Fig. \ref{fig:productstate} (c)(d).
            The partition functions are the products of the weight of the domain wall and the weight of the triangle including the spin $R$, $Z^{(n,k)}_{AB} = W_{DW} W^{R_{\mathbb{I}}}(\mathbb{C},\mathbb{C}; \mathbb{C}) = d^{-\vert \mathbb{C} \vert (L+1)}$  and $Z^{(n,k)}_{AB \cup R} = W_{DW}W^{R_{\mathbb{C}}}(\mathbb{C},\mathbb{C}; \mathbb{C})= d^{-\vert \mathbb{C} \vert L}$, respectively.
            \item Scenario II: the spin $R$ can change the spin configuration with the bulk spins inside its downward light cone fixed to $\mathbb{I}$ as shown in Fig. \ref{fig:productstate} (e)(f) and lead to the formation of a new domain wall. In this case, the partition functions are the weights of the new domain walls, $Z^{(n,k)}_{AB} = W_{DW}  \frac{W^{R_{\mathbb{I}}}(\mathbb{C},\mathbb{C}; \mathbb{I})}{W^{\mathcal{D}}(\mathbb{C},\mathbb{C}; \mathbb{I})} = d^{-\vert \mathbb{C} \vert L}$  and $Z^{(n,k)}_{AB \cup R}  = W_{DW}\frac{W^{R_{\mathbb{C}}}(\mathbb{C},\mathbb{C}; \mathbb{I})}{W^{\mathcal{D}}(\mathbb{C},\mathbb{C}; \mathbb{I})}= d^{-\vert \mathbb{C} \vert (L-1)}$, respectively.
        \end{itemize}
        Therefore, the spin configuration given by Scenario II is dominant, and the mutual information $I_{AB : R} = 2\log (d)$. Here, we have assumed there is no quantum dephasing channel inside the upward light cone of the spin $R$.

        When there is only one quantum dephasing channel inside the upward light cone of the spin $R$, the dominant spin configurations shown in Fig. \ref{fig:productstate} (e)(f) remain unchanged and this quantum dephasing channel can be treated as a local magnetic field. Consequently, the mutual information is $I_{AB : R} = \log (d)$. On the other hand, when there are more than two quantum dephasing channels inside the upward light cone of the spin $R$, the spin configurations shown in Fig. \ref{fig:productstate} (a)(b) are dominant, and the spin $R_{\mathbb{C}}$ can be treated as a local magnetic field. Consequently, the mutual information is $I_{A B: R} = 0$. This is the same as the case with a maximally mixed state as the initial state and thus the dynamics of the mutual information should be collapsed with a rescaled time $t=t_{0}/q^{-1/2}$.
    \end{itemize}
\end{itemize}
 
\begin{figure}[ht]
\centering
\includegraphics[width=0.4\textwidth, keepaspectratio]{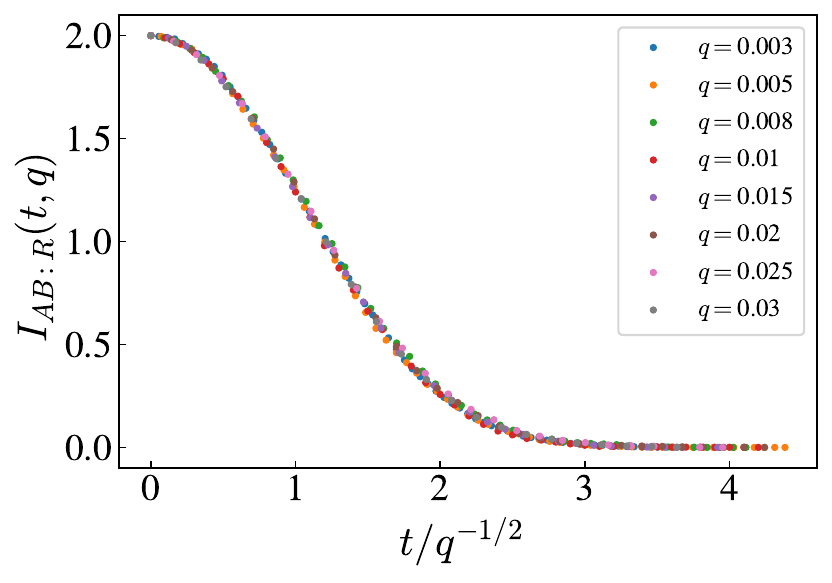}
\caption{Mutual information $I_{AB : R}(t,q)$ vs rescaled time $t/q^{-1/2}$ in the presence of quantum dephasing channels with probability $q$. The system size is $L=512$.}
\label{fig:Dephasing_pm0.0_IAR}
\end{figure}

We have verified the theoretical predictions with the numerical results from the Clifford simulations. The dynamics of the mutual information $I_{AB : R}(t, q)$ can be collapsed with a rescaled time $t/q^{1/2}$ as shown in Fig. \ref{fig:Dephasing_pm0.0_IAR}.

\subsection{Generalization
to other quantum channels}
The preceding discussion has assumed that the quantum noises are modeled by the quantum dephasing channel. The quantum dephasing channel is unital and has no effect on the identity density matrix, i.e., $\mathcal{D}(I)=I$. However, the theoretical prediction of the timescale $q^{-1/2}$ does not depend on the specific choice of the quantum channels, regardless of unital or non-unital channels. We showcase the generalization to the presence of the reset channel that is non-unital below and the extension to other quantum channels is straightforward. 

The density matrix under the action of reset channel on qudit $i$ is 
\begin{eqnarray}
\label{eq:reset}
    \rho^{\prime} = \mathcal{R}_{i}(\rho) = \sum_{j=0}^{d-1} \vert 0 \rangle \langle j \vert \rho \vert j \rangle \langle 0 \vert = \tr_{i} \rho \otimes \vert 0 \rangle \langle 0 \vert_{i}.
\end{eqnarray}
Therefore, when the noise probability is strong enough, e.g., $q=1.0$, the steady state is the product state $\vert 0 \rangle^{\otimes L}$, not a mixed state anymore. In terms of the statistical model, the presence of a reset channel can change the weight of the diagonal bond between two diagonally adjacent spins,
\begin{eqnarray}
    \langle \sigma \vert \mathcal{R} \vert \tau \rangle = d^{r-\vert \tau \vert},
\end{eqnarray}
and thereby affects the weight of the triangle. Similar to the analysis of the quantum dephasing channel shown in Fig. \ref{fig:dephasing_and_bell} (a), we estimate the weights of triangles including a reset channel and with specific spin configurations. 
\begin{itemize}
    \item $(\sigma, \sigma; \sigma)$:
    \begin{eqnarray}
        W^{\mathcal{R}}(\sigma, \sigma; \sigma) &=& \sum_{\tau \in S_{r}} \text{Wg}^{(r)}_{d^{2}} (\sigma^{-1}\tau) d^{r-\vert \sigma^{-1} \tau \vert } \langle \sigma \vert \mathcal{R} \vert \tau \rangle \\ \nonumber
        &=& \sum_{\tau \in S_{r}} \text{Wg}^{(r)}_{d^{2}} (\sigma^{-1}\tau) d^{r-\vert \sigma^{-1} \tau \vert } d^{r-\vert \tau \vert} \\ \nonumber
        &\sim& d^{-\vert \sigma \vert}.
    \end{eqnarray}
    Therefore, it also acts like a magnetic field pinning in the direction $\mathbb{I}$.
    \item  $(\mathbb{C}, \mathbb{C}; \mathbb{I})$:
    \begin{eqnarray}
        W^{\mathcal{R}}(\mathbb{C}, \mathbb{C}; \mathbb{I}) &=&  \sum_{\tau \in S_{r}} \text{Wg}^{(r)}_{d^{2}} (\mathbb{I}^{-1}\tau) d^{r-\vert \mathbb{C}^{-1} \tau \vert } \langle \mathbb{C} \vert \mathcal{R} \vert \tau \rangle \\ \nonumber
        &=& \sum_{\tau \in S_{r}} \text{Wg}^{(r)}_{d^{2}} (\mathbb{I}^{-1}\tau) d^{r-\vert \mathbb{C}^{-1} \tau \vert } d^{r-\vert \tau \vert} \\ \nonumber
        &\sim& d^{- \vert \mathbb{C} \vert}.
    \end{eqnarray}
    The reset channel can also break the unitary constraint but with a larger three-body weight.
\end{itemize}
Now, we discuss the calculation of the free energy:
\begin{itemize}
    \item $F_{R}^{(n,k)}$: the dominant spin configuration is that all the spins are $\mathbb{I}$. The partition function is the weight of the triangle including the spin $R_{\mathbb{C}}$, $Z^{(n,k)}_{R} = W^{R_{\mathbb{C}}}(\mathbb{I}, \mathbb{I}; \mathbb{I}) = d^{-\vert \mathbb{C} \vert}$. Therefore, $F^{(n,k)}_{R} = \vert \mathbb{C} \vert \log (d)$ and thus $S_{R} = \log(d)$.

    \item $F^{(n,k)}_{AB}$ and $F^{(n,k)}_{AB \cup R}$: we also consider the revised hybrid evolution from the top to the bottom to analyze the formation of the domain wall.

    \begin{itemize}
        \item In the absence of the Bell pair, i.e., the spin $R_{\mathbb{I}}$ for $F_{AB}$ or $R_{\mathbb{C}}$ for $F_{AB \cup R}$. If the dominant spin configuration is still to form a domain wall induced by the reset channels near the top boundary such that all other reset channels reside in the region $\mathbb{I}$.
        The total weight is  
        \begin{eqnarray}
            \label{eq:Wdw_reset}
            W_{DW}^{\prime} = W^{0}(\mathbb{C}, \mathbb{I}; \mathbb{I})^{L-2N_{r}}  W^{\mathcal{R}}(\mathbb{C}, \mathbb{C}; \mathbb{I})^{N_{r}} = d^{-\vert \mathbb{C} \vert (L-2N_{r})} d^{-\vert \mathbb{C} \vert N_{r}} = d^{-\vert \mathbb{C} \vert (L-N_{r})},    
        \end{eqnarray}
        which depends on $N_{r}$, the number of reset channels whose light cones contribute to the formation of the domain wall. Therefore, a new deeper domain wall may be formed with some reset channels living in the region $\mathbb{C}$ for a specific trajectory to give the dominant spin configuration. We denote the number of reset channels along the domain wall and living in the region $\mathbb{C}$ as $N^{\prime}_{r}$ and $N^{\prime \prime}_{r}$, respectively. The total weight is 
        \begin{eqnarray}
        W_{DW}^{\prime} W^{\mathcal{R}}(\mathbb{C},\mathbb{C}; \mathbb{C})^{N_{r}^{\prime \prime}} = d^{-\vert \mathbb{C} \vert L + \vert \mathbb{C} \vert (N^{\prime}_{r}-N_{r}^{\prime \prime})},
        \end{eqnarray}
        which is dominant when $N_{r}^{\prime} - N_{r}^{\prime \prime} > N_{r}$. Thus there is a tradeoff in choosing the reset channels to form the domain wall determined by the difference $N^{\prime}_{r}-N_{r}^{\prime \prime}$ for a specific trajectory as shown in Fig. \ref{fig:Reset_competition}. However, on the one hand, the trajectories can realize large $N^{\prime}_{r}$ and small $N_{r}^{\prime \prime}$ are rare; on the other hand, when $N_{r}^{\prime \prime}$ is small, it is reasonable to neglect the effects of these reset channels. Moreover, this contribution from the reset channels residing in the region $\mathbb{C}$ will be canceled in the calculation of $I_{AB : R}$ and $I_{A:B}$ discussed below. Therefore, we can still assume that there are no quantum noises in the region $\mathbb{C}$.

        \begin{figure}[ht]
        \centering
        \includegraphics[width=0.46\textwidth, keepaspectratio]{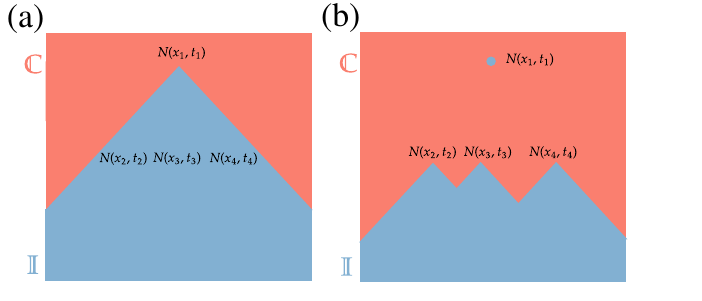}
        \caption{(a) shows the spin configuration with all other reset channels residing in the region $\mathbb{I}$ with $N^{\prime}_{r}=1$ and $N^{\prime \prime}_r=0$. (b) shows the spin configuration with a reset channel residing in the region $\mathbb{C}$ with $N^{\prime}_{r}=3$ and $N^{\prime \prime}_r=1$. Other reset channels in the bulk are not shown here. For this specific trajectory, the spin configuration shown in (b) is dominant.}
        \label{fig:Reset_competition}
        \end{figure}
        
        The difference between the total weights of the domain wall with different quantum channels results in the different steady states: maximally mixed state for quantum dephasing channel and product state for reset channel with $q=1.0$. As shown in Eq. \eqref{eq:Wdw_reset}, when $q=1.0$, $N_{r}=L$ and thus $W^{\prime}_{DW} = d^{0}$. Therefore, $S_{A}=0$ indicating that the steady state is a product state different from the maximally mixed state with quantum dephasing channels. More precisely, when $q=1.0$, we should compare the weights of the triangles in which there are two quantum channels. However, 
        the analysis based on the weights of triangles including one quantum channel is enough to give the correct predictions. 
        \item The spin $R$ is deep (see Fig. \ref{fig:productstate}) (a) and (b)), i.e., far away from the top boundary. In this case, the spin $R_{\mathbb{I}}$ for $F^{(n,k)}_{AB}$ or $R_{\mathbb{C}}$ for $F^{(n,k)}_{AB \cup R}$ must reside in the region $c_{x_{1}, t_{1}} \cup c_{x_{2}, t_{2}} \cup ... \cup c_{x_{N_{r}}, t_{N_{r}}}$. Therefore, the partition functions are the products of the weight of the domain wall and the weight of the triangle including the spin $R$, $Z^{(n,k)}_{AB}=W^{\prime}_{DW}W^{R_{\mathbb{I}}}(\mathbb{I},\mathbb{I}; \mathbb{I}) = W^{\prime}_{DW}$ and  $Z^{n,k}_{AB \cup R} = W^{\prime}_{DW} W^{R_{\mathbb{C}}}(\mathbb{I}, \mathbb{I}; \mathbb{I}) = W^{\prime}_{DW} d^{-\vert \mathbb{C} \vert }$, respectively. Therefore, the mutual information is $I_{AB : R} = 0$, i.e., the encoded information has been destroyed.
        \item  The spin $R$ is shallow (see Fig. \ref{fig:productstate}) (c) and (d)). 
        \begin{itemize}
            \item Scenario I: the spin configurations are shown in Fig. \ref{fig:productstate} (c) and (d). We assume there is no reset channel inside the upward light cone of the spin $R$. The total weight for $F^{(n,k)}_{AB}$ is
            \begin{eqnarray}
                W_{DW}^{\prime}W^{R_{\mathbb{I}}}(\mathbb{C},\mathbb{C};\mathbb{C}) = d^{-\vert \mathbb{C} \vert (L -N_{r} +1) },
            \end{eqnarray}
            and the total weight for $F^{(n,k)}_{AB \cup R}$ is 
            \begin{eqnarray}
                W_{DW}^{\prime}W^{R_{\mathbb{C}}}(\mathbb{C},\mathbb{C};\mathbb{C}) = d^{-\vert \mathbb{C} \vert (L-N_{r})}.
            \end{eqnarray}
            \item Scenario II: the spin configurations are shown in Fig. \ref{fig:productstate} (e) and (f). The total weight for $F^{(n,k)}_{AB}$ is
            \begin{eqnarray}
                W_{DW}^{\prime \prime} \frac{W^{R_{\mathbb{I}}}(\mathbb{C},\mathbb{C};\mathbb{I})}{W^{\mathcal{R}}(\mathbb{C},\mathbb{C};\mathbb{I})} = d^{-\vert \mathbb{C} \vert (L -N_{r} - 1) } \frac{d^{-2\vert \mathbb{C} \vert}}{d^{-\vert \mathbb{C} \vert}} = d^{-\vert \mathbb{C} \vert (L-N_{r})},           
            \end{eqnarray}
            and the total weight for $F^{(n,k)}_{AB \cup R}$ is 
            \begin{eqnarray}
                W_{DW}^{\prime \prime} \frac{W^{R_{\mathbb{C}}}(\mathbb{C},\mathbb{C};\mathbb{I})}{W^{\mathcal{R}}(\mathbb{C},\mathbb{C};\mathbb{I})} = d^{-\vert \mathbb{C} \vert (L -N_{r} - 1) } \frac{d^{-\vert \mathbb{C} \vert}}{d^{-\vert \mathbb{C} \vert}} = d^{-\vert \mathbb{C} \vert (L-N_{r}-1)},          
            \end{eqnarray}
            where we have replaced $W_{DW}^{\prime}$ with $W_{DW}^{\prime \prime}$ because the weight in the presence of reset channels depends on $N_{r}$.
        \end{itemize}
        Therefore, the spin configurations given by Scenario II are dominant. And the mutual information $I_{AB : R} = 2\log(d)$. Based on the same argument, the average dynamics of $I_{AB : R}$ should be collapsed with rescaled time $t/q^{-1/2}$.
    \end{itemize}      
\end{itemize}        

\begin{figure}[ht]
\centering
\includegraphics[width=0.4\textwidth, keepaspectratio]{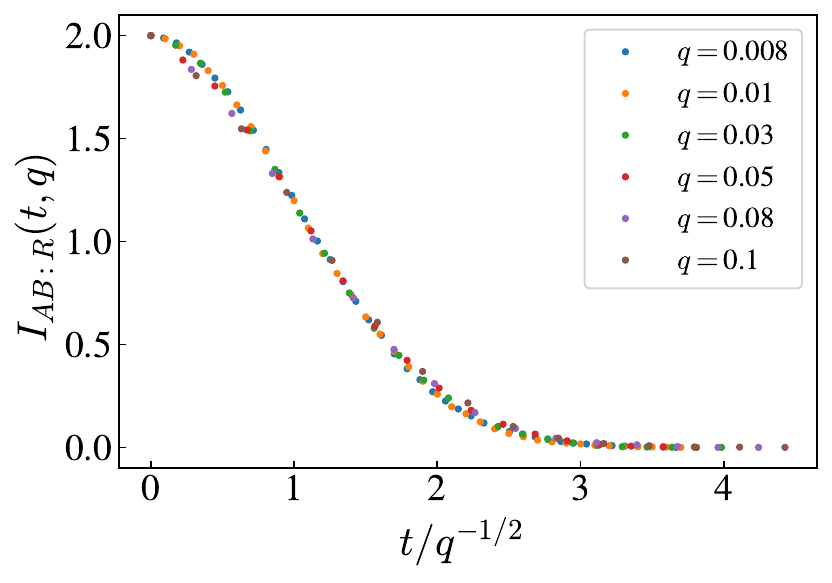}
\caption{Mutual information $I_{AB : R}(t,q)$ vs rescaled time $t/q^{-1/2}$ in the presence of reset quantum channels with probability $q$. The system size is $L=320$.}
\label{fig:Reset_pm0.0_IAR}
\end{figure}

We have verified the theoretical predictions with the numerical results from Clifford simulations. The dynamics of the mutual information $I_{AB : R}(t, q)$ can be collapsed with a rescaled time $t/q^{1/2}$ as shown in Fig. \ref{fig:Reset_pm0.0_IAR}.

\subsection{Generalization to the presence of projective measurements}
In this section, we extend the above analysis to the presence of projective measurements. We can also leave the random locations of measurements as a quenched disorder. However, there should be another copy to calculate the probabilities of measurement outcomes and the degrees of freedom of the statistical model should be formed by the permutation-valued spins that are the elements of the permutation group $S_{r}$ with $r=nk+1$. Specifically,
    \begin{eqnarray}
    \mathbb{C^{\prime}} &=& \mathbb{C}  \otimes \begin{pmatrix} 1 \\ 1 \end{pmatrix} =  \begin{pmatrix} 1 & 2 & ... & n  \\ 2 & 3 & ... & 1\end{pmatrix}^{\otimes k} \otimes \begin{pmatrix} 1 \\ 1 \end{pmatrix},  \\         
    \mathbb{I^{\prime}} &=& \mathbb{I}  \otimes \begin{pmatrix} 1 \\ 1 \end{pmatrix} = \begin{pmatrix} 1 & 2 & ... & n  \\ 1 & 2 & ... & n\end{pmatrix}^{\otimes k} \otimes \begin{pmatrix} 1 \\ 1 \end{pmatrix}.
    \end{eqnarray}
    Due to $\vert \mathbb{C}^{\prime} \vert = \vert \mathbb{C} \vert = k(n-1)$,  $\vert \mathbb{I}^{\prime} \vert = \vert \mathbb{I} \vert = 0$, the unit energy of the domain wall is unchanged. Therefore, we still use $\mathbb{C}$ and $\mathbb{I}$ to represent the degrees of freedom in the effective statistical model.
In the presence of a projective measurement, the weight of the diagonal bond between diagonally adjacent two spins is
\begin{eqnarray}
    w_{d}^{P}(\sigma, \tau) = \langle \sigma \vert P^{\otimes r} \vert \tau \rangle = 1,
\end{eqnarray}
and the three-body weight of the triangle including a projective measurement between $\sigma_{1}$ ad $\tau$ is:
\begin{eqnarray}
    W^{P}(\sigma, \sigma; \sigma) &=& \sum_{\tau \in S_{r}} \text{Wg}_{d^{2}}^{(r)} (\sigma \tau^{-1}) d^{r-\vert \sigma^{-1} \tau \vert} \\ \nonumber 
    & \approx & \sum_{\tau \in S_{r}} \text{Moeb}(\sigma \tau^{-1}) d^{-r-3\vert \sigma^{-1} \tau \vert} \\ \nonumber
    & \approx & d^{-r},
\end{eqnarray}
that is the same as the weight $W^{0}(\sigma, \sigma; \sigma)$ without measurement up to an irrelevant factor $d^{-r}$ appearing in both the numerator and denominator of Eq.~\eqref{eq:smzz}. In the presence of a domain wall, the three-body weight is:
\begin{eqnarray}
    W^{P}(\sigma^{\prime}, \sigma; \sigma) &=& \sum_{\tau \in S_{r}} \text{Wg}_{d^{2}}^{(r)} (\sigma \tau^{-1}) d^{r-\vert \sigma^{-1} \tau \vert} \\ \nonumber 
    & \approx & \sum_{\tau \in S_{r}} \text{Moeb}(\sigma \tau^{-1}) d^{-r-3\vert \sigma^{-1} \tau \vert} \\ \nonumber
    & \approx & d^{-r},
\end{eqnarray}
which indicates the $\sigma^{\prime}$ decouple from the rest two spins. Therefore, the domain wall going through the broken bond with a measurement can gain energy. Consequently, the measurements can be treated as random Gaussian potential, and cause the fluctuation of the domain wall. For the entanglement generation and mutual information scaling with the presence of measurements, please refer to the next subsection.

In terms of the information protection behavior, the fluctuation induced by random measurements does not affect the timescale. This is because the presence of measurements does not quantitatively change the probability that quantum noises appear inside the upward light cone of the encoded information. As the critical height for information protection is in the order of $q^{-1/2}$, the upward light cone from the reference qudit location is well within the $\mathbb{C}$ domain wall with vertical height in the order of $q^{-2/3}$ in the random measurements background. The argument before applied for the information protection timescale thus remains quantitatively the same.

We note that although the problem studied here (with measurements) shares the same effective Hamiltonian as the 2D random bond Ising model for wetting transition~\cite{PhysRevB.39.2632} if the temporally uncorrelated quantum noises are treated as uniform magnetic fields where $q^{-1/2}$ scaling can also be obtained, it neglects the locations of quantum noises which are crucial for $q^{-1/2}$ scaling and would lead to incorrect theoretical predictions for the timescale of information protection in the absence of measurements. 

\subsection{von Neumann entropy and mutual information of the steady state}
In this section, we will give the theoretical analysis of the von Neumann entropy $S_{\alpha}$ and mutual information $I_{A:B}$ within the system at the steady state. 

In the absence of projective measurements, the spin configuration of $S_{AB}$ of the steady state has been depicted in Fig. \ref{fig:productstate} (a), the domain wall has been divided into many smaller domain walls for which the starting and end points are determined by the positions of the quantum channels. 
Therefore, the effective length scale is given by the average distance between two adjacent quantum channels. Although these quantum channels may act at different discrete time steps, we assume they are in the same discrete time step for simplicity. After the average over the randomness of the quantum channels, the effective length scale is:
\begin{eqnarray}
    L_{\rm{eff}} = \sum_{l=1}^{\infty} l (1-q)^{l-1} q = q^{-1},
\end{eqnarray}
where $l$ is the distance between two adjacent quantum channels. As discussed in the last section, the random projective measurements can be treated as Gaussian potential resulting in the fluctuation of the domain wall to cross more bonds with measurements to gain energy. The von Neumann entropy of a subsystem in the absence of quantum noises has been studied before. It can be understood by the KPZ field theory~\cite{PhysRevLett.56.889, PhysRevLett.55.2923, PhysRevLett.55.2924, PhysRevA.39.3053, PhysRevLett.129.080501,Gueudré_2012,barraquandHalfSpaceStationaryKardar2020} which gives:
\begin{eqnarray}
    S_{\alpha} = s_{0}L_{\alpha} + s_{1} L_{\alpha}^{1/3},
\end{eqnarray}
where $L_{\alpha}$ is the size of subsystem $\alpha$, i.e., the distance between the starting point and end point of the domain wall, and $s_{0}$, $s_{1}$ are positive constants. For the case in the presence of temporally correlated quantum noises in the bulk as discussed below or boundary noises~\cite{PhysRevLett.129.080501}, we can also apply KPZ field theory~\cite{PhysRevLett.56.889, PhysRevLett.55.2923, PhysRevLett.55.2924, PhysRevA.39.3053, PhysRevLett.129.080501} straightforwardly to understand the von Neumann entropy with an effective length scale $L_{\text{eff}}$ because the quantum noises just act as emergent boundaries of the domain wall and no other noises encountered in the path of the domain wall. However, for the temporally uncorrelated noises, to understand the $q^{-1/3}$ scaling, we must address why quantum noises with randomly distributed space-time locations do not affect the fluctuation of the domain wall. The answer is provided in this work as discussed above: on the one hand, the effects of quantum noises living in the region $\mathbb{C}$ can be neglected; on the other hand, although there are quantum noises below the domain wall in the absence of projective measurements, these quantum noises can also be neglected because the height of the domain wall predicted by the KPZ field theory with effective length scale $L_{\text{eff}} \sim q^{-1}$ is $q^{-2/3}$ in the presence of measurements, which is smaller than the original height $q^{-1}$. Therefore, the effects of temporally uncorrelated quantum noises can be attributed to inducing an effective length scale.

Now, we show how to get the analytical predictions of von Neumann entropy and mutual information of the steady state. The probability of projective measurement is set to $p_{m} < p_{m}^{c}$. 
\begin{itemize}
    \item von Neumann entropy $S_{\alpha}$:
    Based on the KPZ field theory, the free energy density with length scale $l$ is
    \begin{eqnarray}
    f(l) &=& F(l)/l \\ \nonumber 
         &=& (s_{0}l+s_{1}l^{1/3}) / l \\ \nonumber
         &=& (s_{0} + s_{1}l^{-2/3}).
    \end{eqnarray}
    Thus the free energy density averaged over the random quantum noises is:
    \begin{eqnarray}
        \langle f(q) \rangle &=& \sum_{l=1}^{\infty} (s_{0}+s_{1}l^{-2/3}) (1-q)^{l-1}q \\ \nonumber
        &\approx& s_{0} + s_{1} (q^{2/3} \Gamma [1/3] + q \zeta [2/3] ) ,
    \end{eqnarray}
    where $\Gamma$ and $\zeta$ are Gamma and Zeta functions respectively and we have neglected the higher-order terms above $q^{1}$. $\Gamma[1/3] \approx 2.68 > 0$ and $\zeta[2/3] \approx -2.45 < 0$. Therefore, the total free energy, i.e., von Neumann entropy, reads:
    \begin{eqnarray}
        \langle F_{\alpha}(q) \rangle \approx s_{0}L_{\alpha} + s_{1} L_{\alpha}(q^{2/3} \Gamma [1/3] + q \zeta [2/3] ),
    \label{eq:f}
    \end{eqnarray}
    where $L_{\alpha}$ is the size of subsystem $\alpha$. The numerical results from Clifford simulations are shown in Fig. \ref{fig:Reset_pm0.2_SASAB} and Fig. \ref{fig:Dephasing_pm0.2_SASAB} for reset channels and quantum dephasing channels, respectively. The signs of the coefficients obtained from the fitting are consistent with the theoretical predictions.

    \begin{figure}[ht]
    \centering
    \includegraphics[width=0.8\textwidth, keepaspectratio]{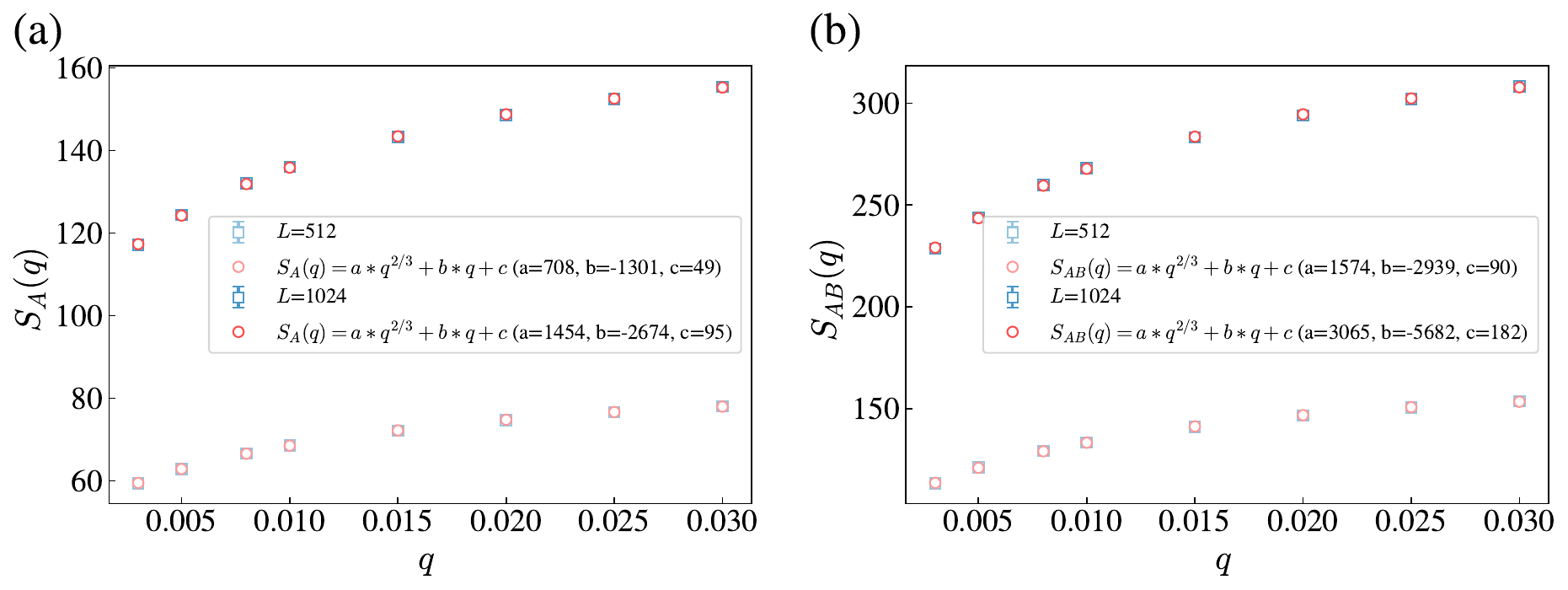}
    \caption{The probability of reset channels is $q$. (a) shows the von Neumann entropy $S_{A}$ and (b) shows the von Neumann entropy $S_{AB}$. The fitting function is chosen as that in Eq. \eqref{eq:f} and the signs of the coefficient are consistent. The probability of projective measurement is $p_{m}=0.2<p_{m}^{c}$.}
    \label{fig:Reset_pm0.2_SASAB}
    \end{figure}

    \begin{figure}[ht]
    \centering
    \includegraphics[width=0.8\textwidth, keepaspectratio]{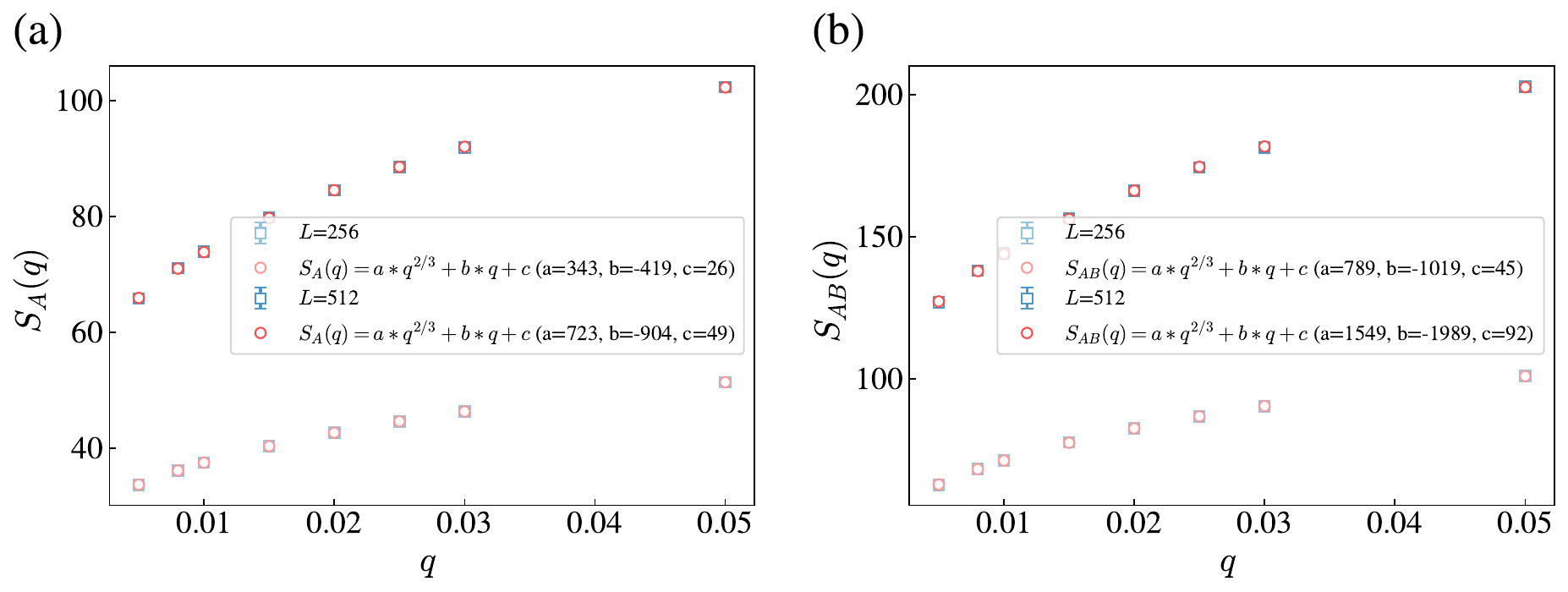}
    \caption{The probability of quantum dephasing channels is $q$. (a) shows the von Neumann entropy $S_{A}$ and (b) shows the von Neumann entropy $S_{AB}$. The fitting function is chosen as that in Eq. \eqref{eq:f} and the signs of the coefficient are consistent. The probability of projective measurement is $p_{m}=0.2<p_{m}^{c}$.}
    \label{fig:Dephasing_pm0.2_SASAB}
    \end{figure}
    
    \item Mutual information $I_{A:B}$: the mutual information is given by:
    \begin{eqnarray}
        I_{A:B} = S_{A}+S_{B}-S_{AB}.
    \end{eqnarray}
    If we replace $S_{\alpha}$ with Eq. \eqref{eq:f}, it seems like $I_{A:B}=0.0$. We will show that the mutual information is given by the contribution of the boundary term of $S_{\alpha}$ which has been neglected in the calculation of Eq. \eqref{eq:f}.

    For the domain walls divided by the quantum noises that do not cross the midpoint $x_{m}$ of the system, the contribution from $S_{A}$, $S_{B}$, and $S_{AB}$ will cancel each other. However, for the domain wall in $S_{AB}$ that crosses the midpoint with the starting point and end point denoted as $x_{0}$ and $x_{1}$, its starting and end points are $(x_{0}, x_{m})$ and $(x_{m}, x_{1})$ for $S_{A}$ and $S_{B}$ respectively. Therefore, this boundary contribution can not be canceled because of 
    \begin{eqnarray}
        F_{AB} = (s_{0}l+s_{1}l^{1/3}),
    \end{eqnarray}
    where $l$ is the length scale of the domain wall which crosses the midpoint and 
    \begin{eqnarray}
        F_{A} + F_{B} &=& s_{0} + s_{1} (l_{A}^{1/3} + (l-l_{A})^{1/3}) \\ \nonumber
        &=& s_{0} + s_{1}l^{1/3} (r^{1/3}+(1-r)^{1/3}),
    \end{eqnarray}
    where we assume $l_{A}=rl$ with $r \in [0,1]$. Consequently, this boundary term contribution to the mutual information after the disorder average is:
    \begin{eqnarray}
        \overline{\langle F_{A}+F_{B}-F_{AB} \rangle} &=&  \overline{\sum_{l=1}^{\infty} (s_{1}l^{1/3}(r^{1/3}+(1-r)^{1/3}-1)) (1-q)^{l-1}q } \\ \nonumber
        &\approx& \overline{s_{1}(r^{1/3}+(1-r)^{1/3}-1}) (\Gamma[4/3]q^{-1/3} + \frac{\Gamma[4/3]}{3}q^{2/3} + \zeta[-1/3]q) \\ \nonumber
        &=& \frac{s_{1}}{2} (\Gamma[4/3]q^{-1/3} + \frac{\Gamma[4/3]}{3}q^{2/3} + \zeta[-1/3]q) \\ \nonumber 
        &\approx&  \frac{s_{1}\Gamma[4/3]}{2}q^{-1/3} \ (q \ll 1.0),
    \end{eqnarray}
    which gives the $q^{-1/3}$ scaling discussed in the main text. The additional numerical results in the presence of quantum dephasing channels are shown in Fig. \ref{fig:Dephasing_pm0.2_IABIAR}.

    \begin{figure}[ht]
    \centering
    \includegraphics[width=0.6\textwidth, keepaspectratio]{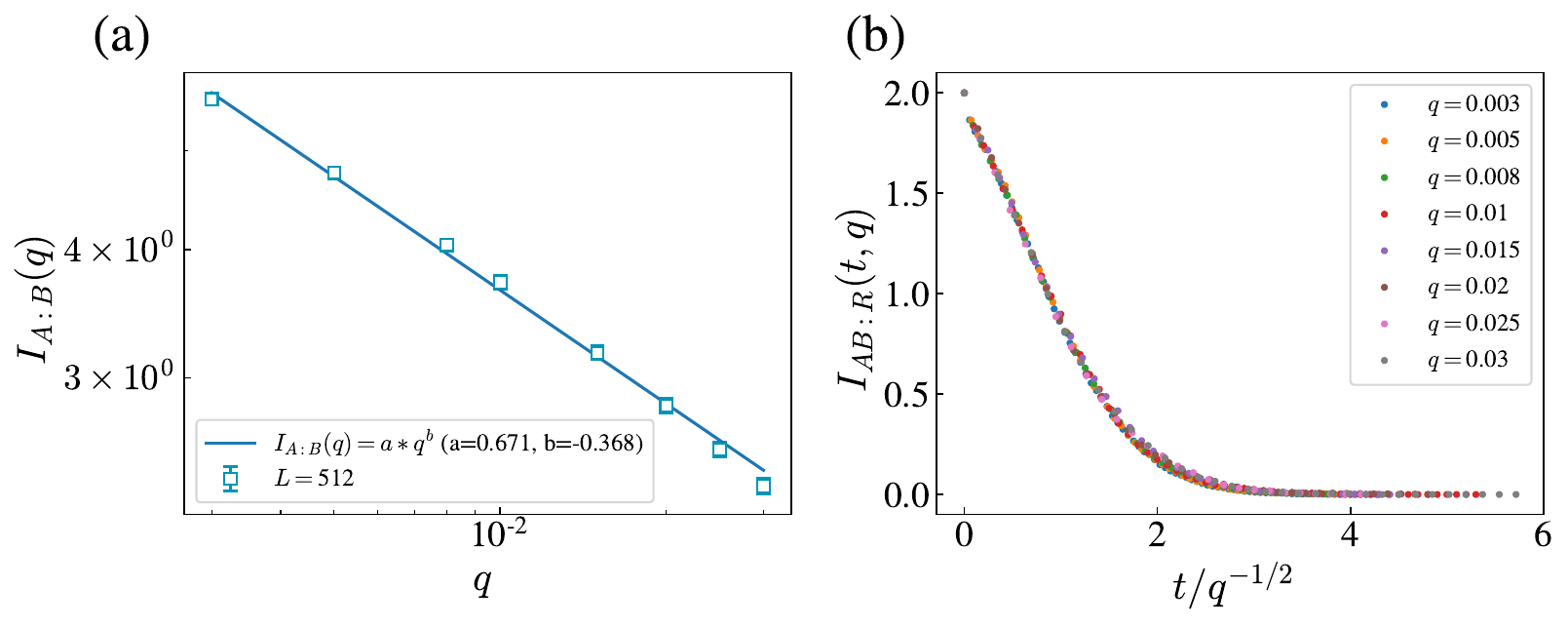}
    \caption{The probability of the quantum dephasing channel is $q$. The system size is $L=512$ and the probability of measurement is $p_{m}=0.2<p_{m}^{c}$. (a) shows the mutual information of the steady state $I_{A:B}$ vs the noise probability $q$; (b) shows the dynamics $I_{AB : R}$ vs rescaled time $t/q^{-1/2}$.}
    \label{fig:Dephasing_pm0.2_IABIAR}
    \end{figure}

\end{itemize}

\section{The analytical understanding in the presence of temporally correlated quantum noises}
In this section, we show the theoretical analysis of the scaling of mutual information and the timescale of information protection in the presence of temporally correlated quantum noise.

\subsection{von Neumann entropy and mutual information of the steady state}
\begin{figure}[ht]
\centering
\includegraphics[width=0.8\textwidth, keepaspectratio]{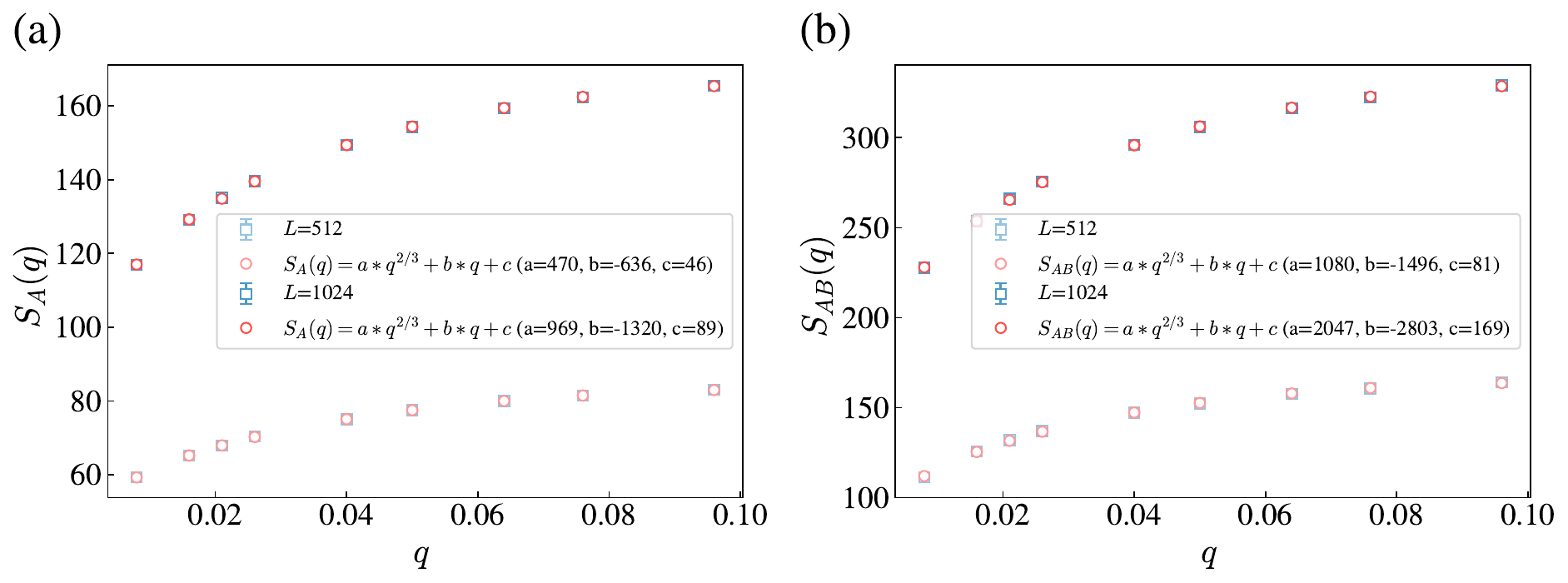}
\caption{The probability of temporally correlated reset channels is $q$. (a) shows the von Neumann entropy $S_{A}$ and (b) shows the von Neumann entropy $S_{AB}$. The fitting function is chosen as that in Eq. \eqref{eq:f} and the signs of the coefficient are consistent with the theoretical predictions. The probability of projective measurement is $p_{m}=0.2<p_{m}^{c}$.}
\label{fig:Reset_pm0.2_SASAB_static}
\end{figure}

\begin{figure}[ht]
\centering
\includegraphics[width=0.6\textwidth, keepaspectratio]{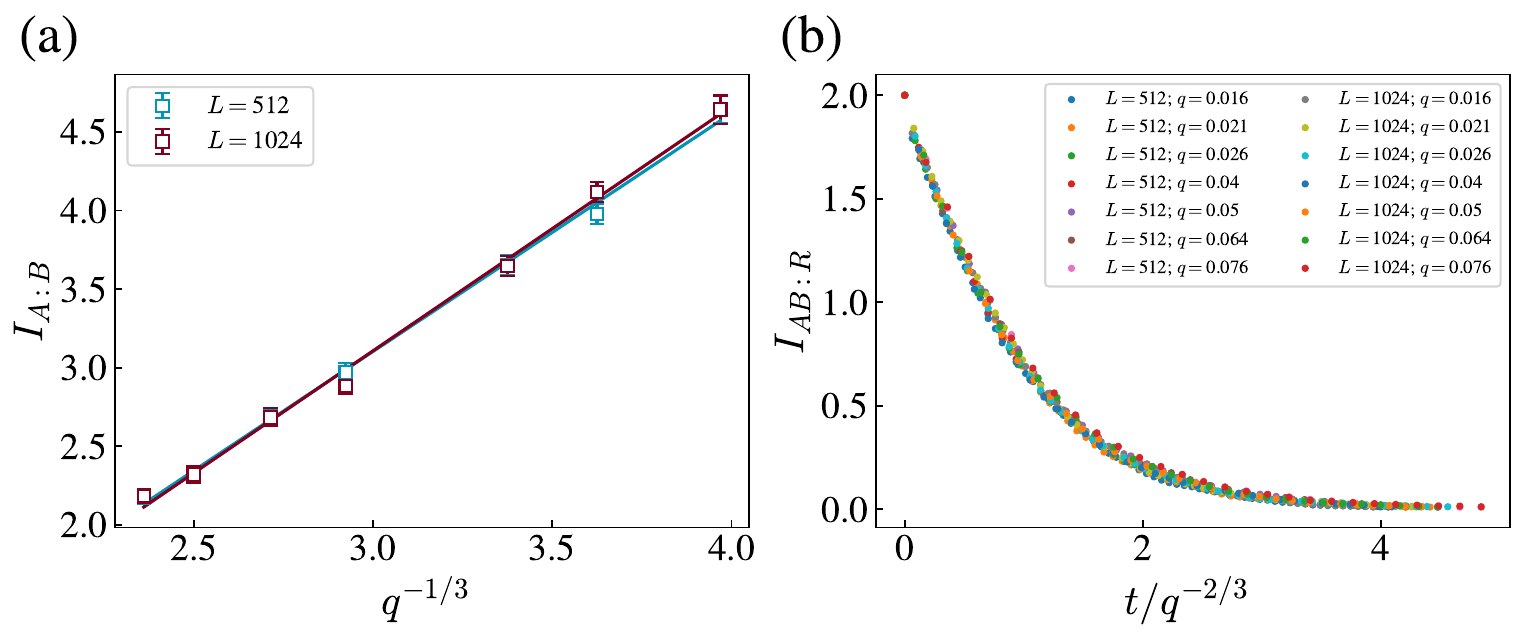}
\caption{The probability of temporally correlated reset channels is $q$ and the probability of projective measurement is $p_{m}=0.2<p_{m}^{c}$. (a) shows the mutual information $I_{A:B}$ vs noise probability $q$ and (b) shows the dynamics of $I_{AB : R}$ vs rescaled time $t/q^{-2/3}$. The timescale of information protection with temporally correlated quantum noises is larger than that with uncorrelated quantum noises.}
\label{fig:Reset_pm0.2_IABIAR_static}
\end{figure}

The temporally correlated quantum noises show a stripe pattern along the time direction and can thus be treated as the emergent boundaries with fixed spins $\mathbb{I}$. This picture leads to the same effective length scale $L_{\text{eff}} \sim q^{-1}$ as in the Markovian noise case. Within adjacent emergent boundaries, there are no other quantum noises, and the free energy of the domain wall is given by the prediction of KPZ field theory and the contribution from the boundary term leads to the $q^{-1/3}$ scaling of mutual information for entanglement generation. The numerical results of the von Neumann entropy are shown in Fig. \ref{fig:Reset_pm0.2_SASAB_static} and the numerical results of the mutual information are shown in the main text and Fig. \ref{fig:Reset_pm0.2_IABIAR_static} (a).

\subsection{The timescale of the information protection}
\begin{figure}[ht]
\centering
\includegraphics[width=0.6\textwidth, keepaspectratio]{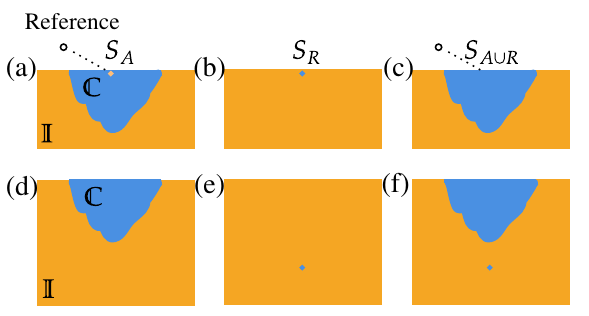}
\caption{The connection between the $L_{\text{eff}}^{2/3}$ timescale of information protection and the height of the domain wall. When the spin $R$ is shallow as shown in (a)(b)(c), the spin $R_{\mathbb{I}}$ in region $\mathbb{C}$ (a) and the spin $R_{\mathbb{C}}$ in region $\mathbb{I}$ (b) act as local magnetic fields. The information is protected. When the spin $R$ is deep as shown in (d)(e)(f), the spin $R_{\mathbb{C}}$ in region $\mathbb{I}$ (e) and the spin $R_{\mathbb{C}}$ in region $\mathbb{I}$ (f) act as local magnetic fields. The information is destroyed.}
\label{fig:MIPT1}
\end{figure}

In the presence of temporally correlated quantum noises, the timescale of information protection will change because there is no quantum noise inside the upward light cone of the spin $R$. The timescale of information protection will be determined by the height of the domain wall which is $L_{\text{eff}}^{2/3}$ from KPZ field theory. We have depicted the theoretical analysis of $L_{\text{eff}}^{2/3}$ timescale in terms of the effective statistical model in Fig. \ref{fig:MIPT1}. Different from the cases with temporally uncorrelated quantum noises, the timescales of information protection in the presence of temporally correlated quantum noises with and without measurements are different. The former is $q^{-2/3}$ while the latter is $q^{-1}$. Therefore, the steady state information protection setting proposed in this works highlights the distinctions between the effects of temporally correlated and uncorrelated quantum noises although they both induce the $q^{-1/3}$ scaling in entanglement generation. The timescale of information protection with temporally correlated quantum noises is larger than that with temporally uncorrelated quantum noises, reflecting the potential benefits of non-Markovianity in quantum dynamics, consistent with recent studies.

\begin{figure}[ht]
\centering
\includegraphics[width=0.4\textwidth, keepaspectratio]{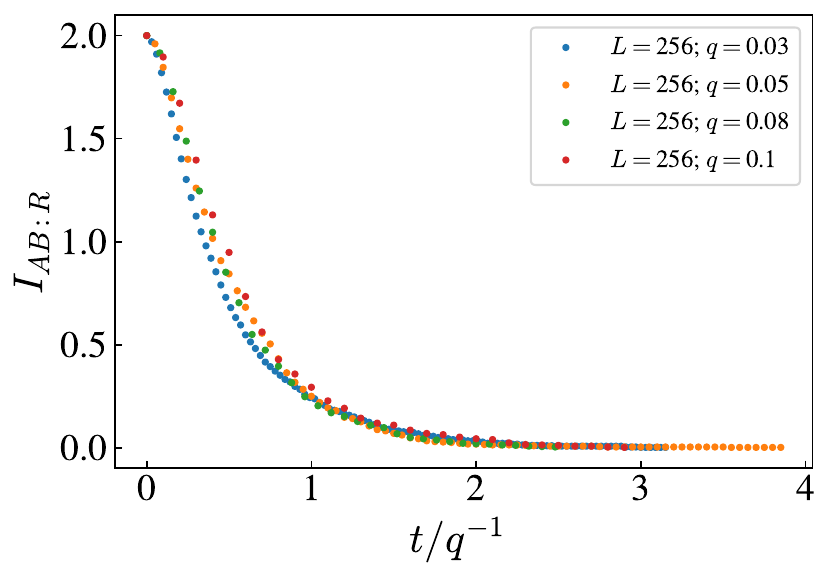}
\caption{The probability of temporally correlated dephasing channels is $q$ and the probability of projective measurement is $p_{m}=0.0$. The dynamics of $I_{AB : R}$ vs rescaled time $t/q^{-1}$. The timescale of information protection with temporally correlated quantum noises corresponds to the average height of the domain wall which is $q^{-1}$ without measurements.}
\label{fig:Dephasing_pm0.0_IAR_static}
\end{figure}

The dynamics of mutual information $I_{AB : R}$ with and without measurements are shown in Fig. \ref{fig:Reset_pm0.2_IABIAR_static} (b) and Fig. \ref{fig:Dephasing_pm0.0_IAR_static}, respectively.

\section{Numerical results for larger measurement rate $p_{m}>p_{m}^{c}$}
\begin{figure}[ht]
    \centering
    \includegraphics[width=0.65\textwidth, keepaspectratio]{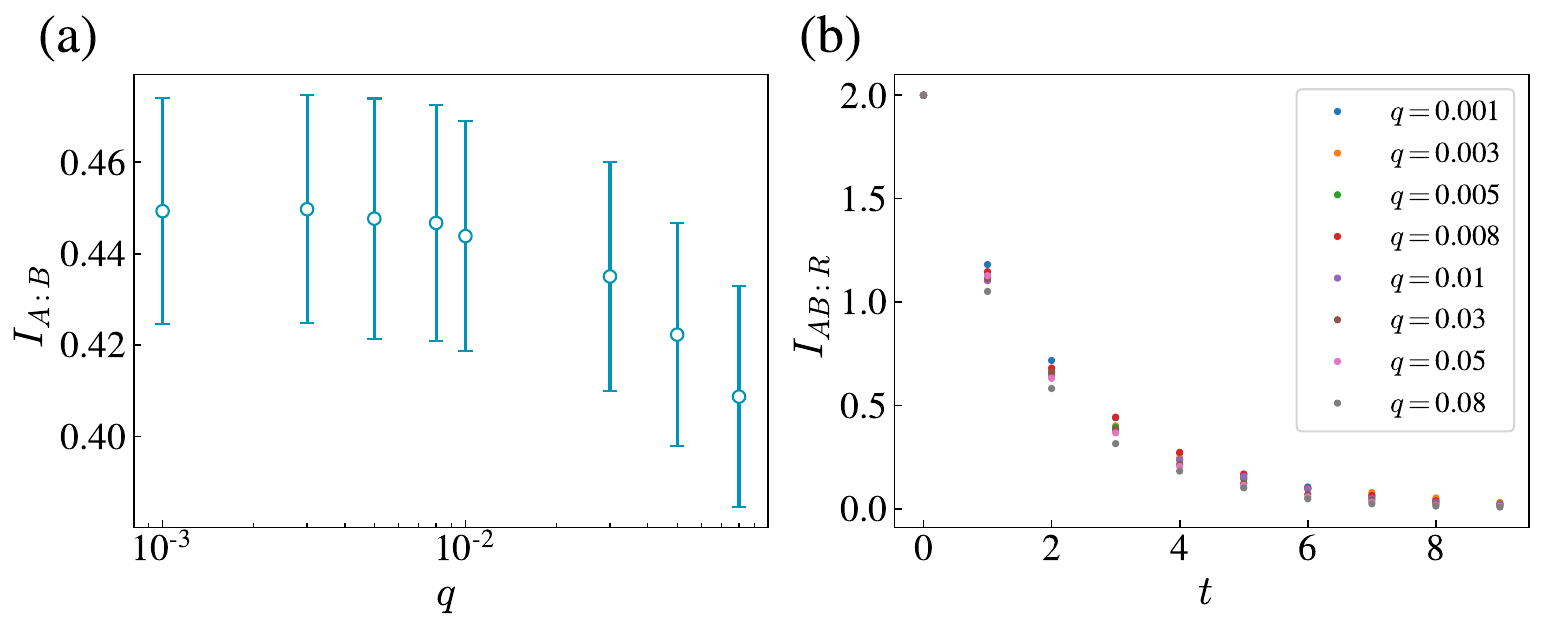}
    \caption{Entanglement generation and information protection for measurement rate larger than the critical value. 
    The probability of projective measurement is $p_{m}=0.6>p_{m}^{c}$ and the probability of temporally uncorrelated quantum noise modeled by reset channel is $q$. (a) shows the mutual information within the system $I_{A:B}$ vs noise probability $q$; (b) shows the dynamics of the mutual information between the system and the reference qubit after the system reaches the steady state. Here, we set system size $L=320$.}
    \label{fig:pm0.6}
    \end{figure}

In the main text, we have provided a comprehensive theoretical analysis of the effects of quantum noises with different types and different space-time distributions on the entanglement generation and information protection when the measurement rate is below the critical rate $p_{m}^{c}$ of MIPT. 
In the absence of quantum noises, the system enters a disordered phase when the measurement rate is larger than $p_{m}^{c}$, because there are so many broken bonds, and thus the crossing of a domain wall over these broken bonds does not cost extra energy. 
We have also numerically investigated the entanglement behaviors and information protection for the case with $p_{m}>p_{m}^{c}$. 
The numerical results are shown in Fig.~\ref{fig:pm0.6} where the temporally uncorrelated quantum noises are modeled by reset channels. 
In the presence of quantum noises with small probability, we find the entanglement within the system is the same as that without noises. 
In $q \to 1$ limit, the steady state is totally determined by the quantum noises. 
Therefore, there is no correlation between any qubits and the rest qubits in the system. 
Consequently, the entanglement decays to zero. 
In the intermediate region, quantum noises simply reduce the area-law amount of entanglement. 
Moreover, from the perspective of information protection, the timescale changes from $q^{-1/2}$ ($p_{m}<p_{m}^{c}$) to a $q$-independent constant for small $q$ ($p_{m}>p_{m}^{c}$), consistent with the fact that the entanglement does not depend on noise probability $q$ when $q$ is small.

\section{Information protection with other setups}
\subsection{Initial state information protection protocol}
In the main text, we have discussed the steady state information protection where the information is encoded into the steady state. A natural question arises as to whether the information protection timescale depends on the state to encode information. A relevant setup is to encode information into the initial state~\cite{Coding_Vijay, liu2024noise}. The state of the system and the reference qubit is $\vert 0 \rangle^{\otimes L-1} \otimes \frac{\vert 00 \rangle_{ER} + \vert 11 \rangle_{ER}}{2}$, where $E$ represents the qubit coupled with the reference qubit, and evolves by the noisy hybrid quantum circuits. The information protection timescale is $O(q^{-1})$ for both temporally correlated and uncorrelated quantum noises, arising from the competition between two spin configurations corresponding to $S_{AB}$ or $S_{AB \cup R}$: one is the spin configuration where all the spins are $\mathbb{C}$ with free energy $O(qLt)$ corresponding to that the information remains in the system; the other is the spin configuration with a domain wall with a leading term of free energy $O(L)$ corresponding to the information leaked into the environment. Therefore, the information protection timescales indeed depend on the state to encode information. Moreover, compared to the steady state information protection protocol investigated in the main text, the initial state information protection cannot distinguish the difference between temporally correlated and uncorrelated noises as well as the entanglement perspective.

\subsection{Information protection without noises}
\begin{figure}[ht]
\centering
\includegraphics[width=0.6\textwidth, keepaspectratio]{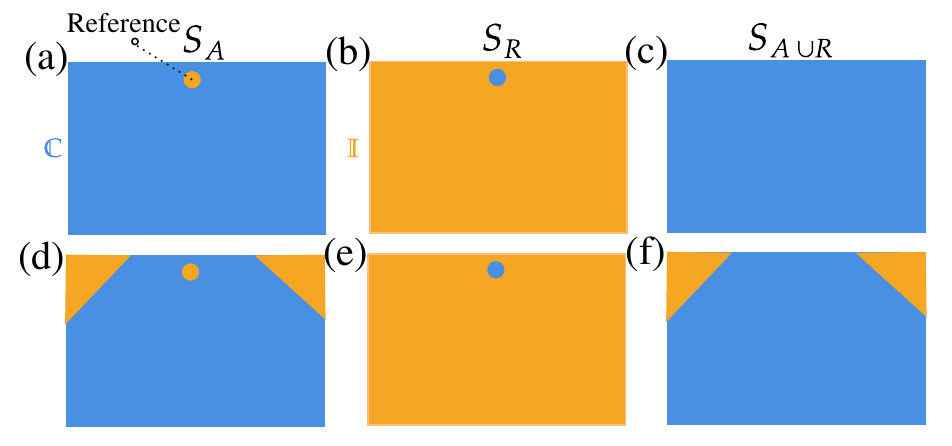}
\caption{Top panel shows the dominant spin configurations with $L_{A}=L$, the limit case $L_{A}>L/2$, and the bottom panel shows the dominant spin configurations with $L_{A}>L/2$. When the subsystem size $L_{A}>L/2$, the encoded information can be perfectly protected.}
\label{fig:MIPT2}
\end{figure}

\begin{figure}[ht]
\centering
\includegraphics[width=0.38\textwidth, keepaspectratio]{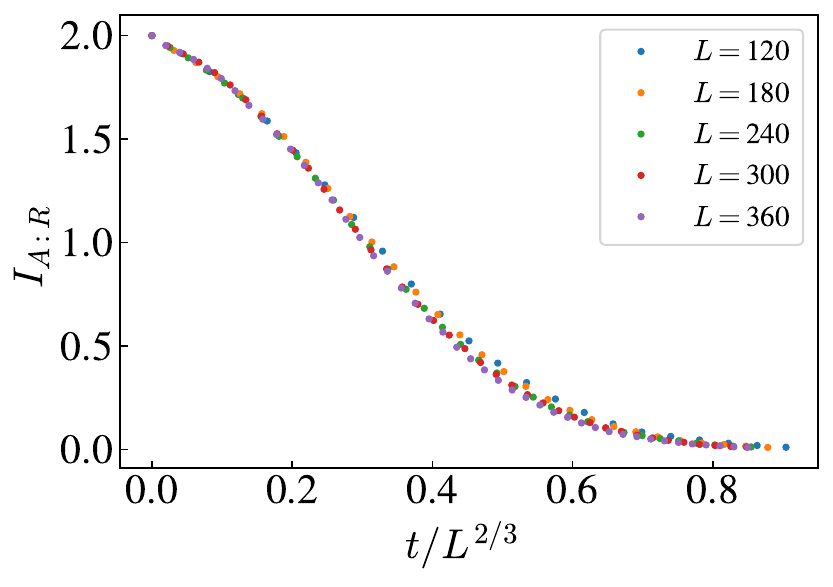}
\caption{MIPT setup with the probability of measurement is $p_{m}=0.1$. The dynamics of mutual information between subsystem $A=[L/3,2L/3]$ and the ancilla qubit can be collapsed with rescaled time $t/L^{2/3}$.}
\label{fig:Mipt_pm0.1_IAR}
\end{figure}

In this section, we will discuss the abilities of the information protection of the noiseless hybrid quantum circuits. For the MIPT setup, when the measurement probability is lower than the critical probability, the von Neumann entropy within the system obeys volume-law, allowing for the perfect protection of the encoded information. This conclusion can also be easily understood from the effective statistical model as shown in Fig. \ref{fig:MIPT2}. However, if we consider a part of the system ($\bar{A}$) as the environment, the timescale of information protection of the remaining subsystem ($A$) will depend on the subsystem size of $A$ denoted as $L_{A}$. Here, $A$ includes the qudit coupled with the reference qudit. If $L_{A}<L/2$, this is similar to the cases with temporally correlated quantum noises by replacing $L_{\text{eff}}$ with $L_{A}$ as shown in Fig. \ref{fig:MIPT1} and thus the timescale of information protection is $L_{A}^{2/3}$. The numerical results are shown in Fig. \ref{fig:Mipt_pm0.1_IAR}. On the other hand, if $L_{A}>L/2$, other spin configurations as shown in Fig. \ref{fig:MIPT2} dominate and the information can still be perfectly protected. We can also consider the initial state information protection protocol in the noiseless case, where the information protection protocol is $L_{A}$ and $\infty$ for $L_{A} < L/2$ and $L_{A} > L/2$ respectively~\cite{subsystemcapacity}.

We have summarized the results of information protection timescales with different setups in Table. \ref{tab:1}.

\begin{table}[]
\caption{Information protection timescales with different encoding schemes ($0<p_{m}<p_{m}^{c}$).\\}\label{tab:1}
\begin{tabular}{@{} p{4cm}p{6cm}p{6cm} @{}}
\toprule
                       &\parbox{5.5cm}{product state (initial state)} & \parbox{6cm}{\textcolor{magenta}{steady state}} \\ 
\midrule
subsystem ($L_{A}<L/2$) without noises    &   \parbox{5.5cm}{$O(L_{A})$}    &     \parbox{5.5cm}{\textcolor{magenta}{$O(L_{A}^{2/3}$}) } 
\\ &&
\\
whole system with temporally correlated noises &           \parbox{5.5cm}{$O(q^{-1})$}             &    \parbox{5.5cm}{\textcolor{magenta}{$O(q^{-2/3})$ }}       \\
&& \\
whole system with temporally uncorrelated noises &           \parbox{5.5cm}{$O(q^{-1})$}             &    \parbox{5.5cm}{\textcolor{magenta}{$O(q^{-1/2})$ }}        \\ \bottomrule
\end{tabular}
\end{table}

\section{Connection between information protection and temporal correlation of quantum noises}
\begin{figure}[ht]
\centering
\includegraphics[width=0.7\textwidth, keepaspectratio]{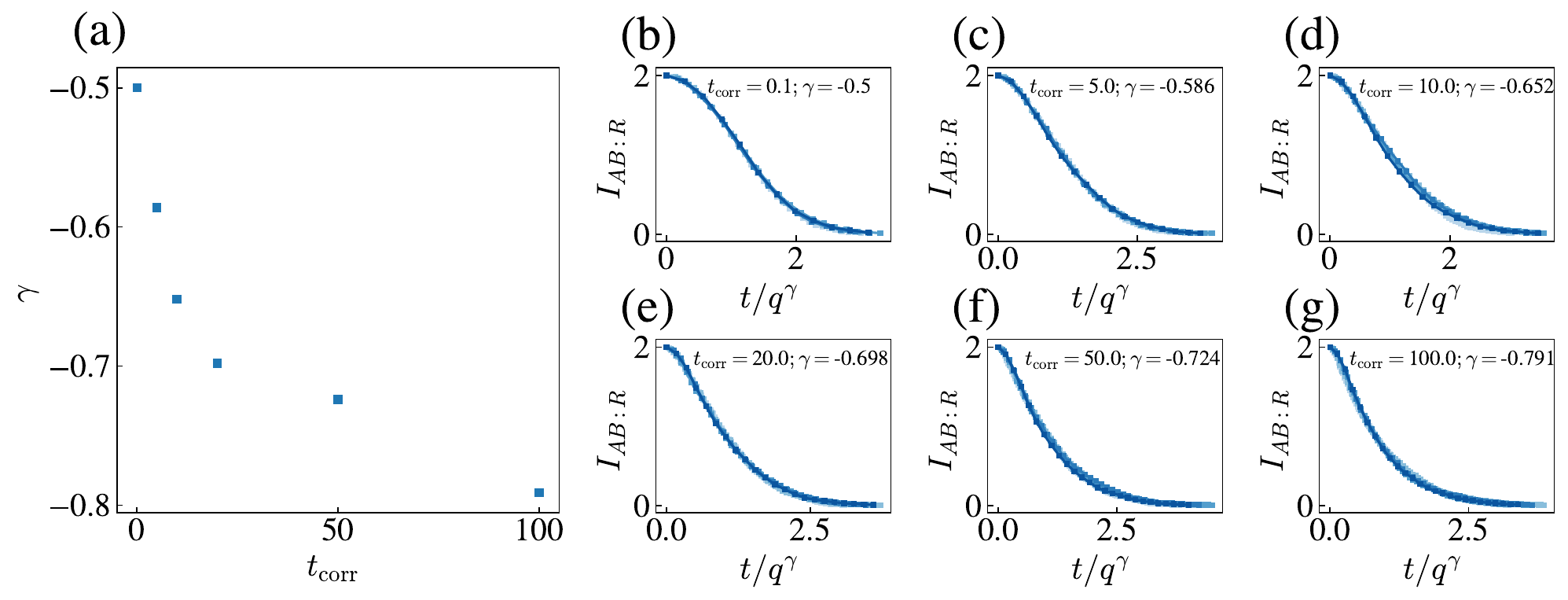}
\caption{Information protection in the presence of general non-Markovian noise with different $t_{\text{corr}}$: (a) shows $\gamma$ vs $t_{\text{corr}}$. (b-g) show the data collapse for different $t_{\text{corr}}$. Here, $L=512$, $q=[0.01,0.08]$.}
\label{fig:constant_tcorr}
\end{figure}

\begin{figure}[ht]
\centering
\includegraphics[width=0.7\textwidth, keepaspectratio]{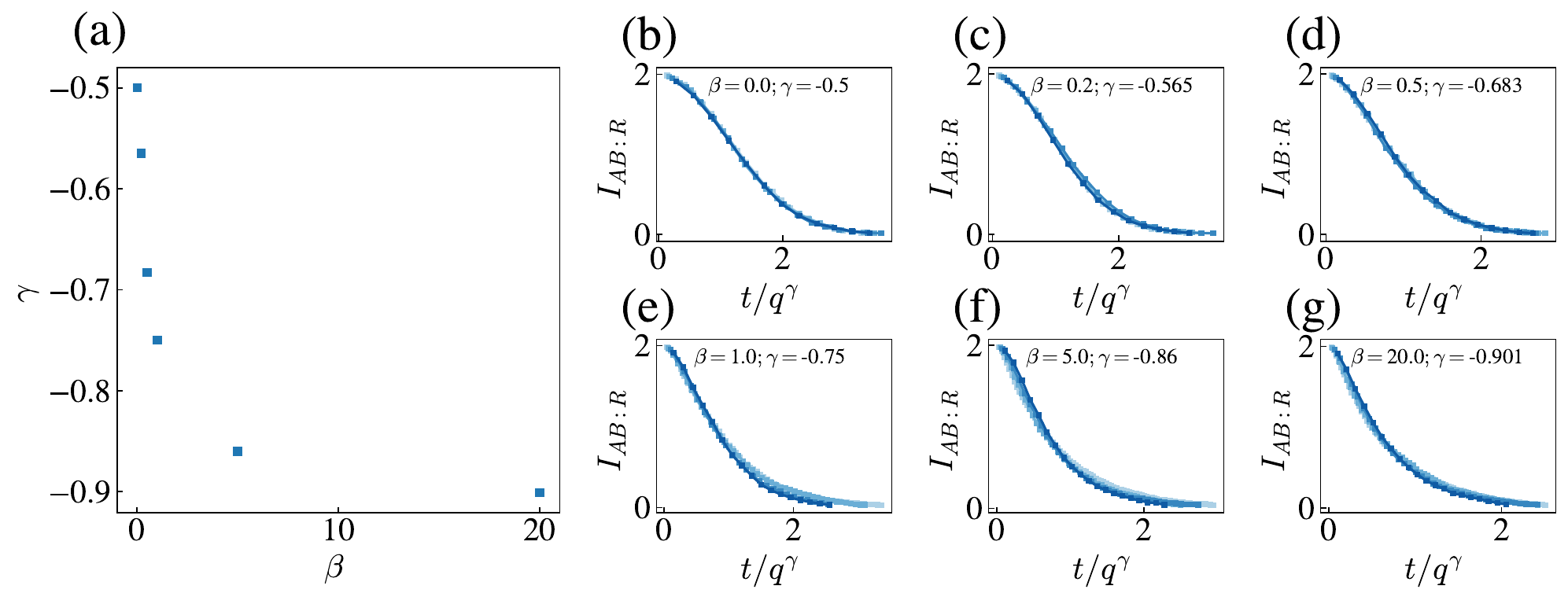}
\caption{Information protection in the presence of general non-Markovian noise with noise strength dependent correlation ($t_\text{corr} = q^{-\beta}$): (a) shows $\gamma$ vs $\beta$. (b-g) show the data collapse for different $\beta$. Here, $L=256$, $q=[0.02,0.08]$.}
\label{fig:qdependent_tcorr}
\end{figure}

As discussed above, in the absence of projective measurements, the timescale of information protection is $q^{-1/2}$ and $q^{-1}$ for uncorrelated noises (Markovian limit) and correlated noises (strong non-Markovian limit) respectively. 
A natural question arises as to the behavior of information protection timescale for a general non-Markovian noise between the Markovian limit and the strong non-Markovian limit.
More specifically, by tuning the correlation length of quantum noise, does there exist a phase transition or crossover in the scaling of the information protection timescale? 
To answer this question, we investigate the effect of correlated noises which satisfies
\begin{eqnarray}
    \langle q_{t}q_{t+\Delta t} \rangle - \langle q_{t} \rangle \langle q_{t+\Delta t} \rangle = q(1-q) e^{-\Delta t / t_{\text{corr}}},
\end{eqnarray}
where $q_{t}$ is a discrete random variable $0$ or $1$ at time step $t$ with average probability $\langle q_{t} \rangle = q$~\cite{Kattemolle2023_z}. $q_t=1$ indicates that there is a noise channel in the time slice $t$. 
$t_{\text{corr}}$ is the correlation timescale of the quantum noise, where $t_{\text{corr}} = 0$ and  $t_{\text{corr}} >0$ corresponds to the Markovian limit and a generic non-Markovian case, respectively. 
The numerical results shown in Fig. \ref{fig:constant_tcorr} with $q$-independent time correlation and Fig. \ref{fig:qdependent_tcorr} with $q$-dependent time correlation ($t_\text{corr} = q^{-\beta}$) both suggest that a crossover occurs, with the scaling of the information protection timescale $q^{\gamma}$ changing from $q^{-1/2}$ to $q^{-1}$ by tuning $t_{\text{corr}}$.

\section{Details of numerical simulation}

	In this section, we introduce more details of numerical simulation. We utilize stabilizer formalism to perform numerical simulations to validate the theoretical predictions in the main text. For a special class of quantum states, we can use a set of Pauli strings, denoted as the stabilizer group $\mathcal{G}$, to uniquely identify a quantum state. The density matrix $\rho$ of the stabilizer state is 
    \begin{eqnarray}
        \rho = \frac{2^{\vert \mathcal{G} \vert}}{2^{L}} \prod_{i=1}^{\vert \mathcal{G} \vert} (\frac{I+g_{i}}{2}),
    \end{eqnarray}
    where $L$ is the system size, $g_{i}$ is $i$-th generator of the stabilizer group and $\vert \mathcal{G} \vert$ is the dimension of group $\mathcal{G}$. The initial state is chosen as $\vert 00...0\rangle$ and the corresponding stabilizer group is $\{Z_{0}, Z_{1}, ..., Z_{L-1} \}$, where $Z_{i}$ represents the Pauli string with Pauli-Z operator on $i$-th qubit and identity operators on other qubits. Each gate is independently drawn from 2-qubit Clifford ensemble. The dynamics of the density matrix can be efficiently simulated classically by keeping track of the evolution of the generators of the stabilizer group: the generators change from $\{g_{i},...,g_{\vert \mathcal{G} \vert} \}$ to $\{g^{\prime}_{i},...,g^{\prime}_{\vert \mathcal{G} \vert} \}$ under the evolution of Clifford gate. Therefore, the computation resources are $O(L^{2})$ instead of $O(2^{L})$.
    
    We can implement single qubit projective measurement $(h=Z_{i})$ in a computational basis as follows: if $h$ commutes with all generators of $\mathcal{G}$ without being an element of $\mathcal{G}$ itself (which is possible for mixed state with $\vert \mathcal{G} \vert < L$), $\pm h$ is simply added as a new generator with equal probabilities; if $h$ commutes with all generators of $\mathcal{G}$ and $\vert \mathcal{G} \vert = L$, the generators stay unchanged; if $h$ anticommutes with a generator $g_{l}$, this generator $g_{l}$ is replaced with $\pm h$ with equal probabilities; if more than one generator anticommute with $h$, we can choose on generator, denoted as $g_{l}$, and multiply the other generators that anticommute with $h$ by $g_{l}$ so that only one generator anticommutes with $h$. 

    We can implement the reset channel $\mathcal{R}_{i}$ as shown in Eq.~\ref{eq:reset} by operating a swap operator between the $i$-th qubit and an ancilla qubit initialized to the state $\vert 0 \rangle$. The density matrix of the system after the reset channel can be obtained by tracing out the ancilla qubit. As shown above, there is at most one generator $g_{l}$ that anticommutes with $Z_{i}$. Therefore, we can implement the quantum dephasing channel $\mathcal{D}_{i}$ shown in Eq.~\ref{eq:dephasing} by eliminating $g_{l}$ from $\mathcal{G}$. If all generators of $\mathcal{G}$ commute with $Z_{i}$, then the generators of $\mathcal{G}$ stay unchanged.

    To create a Bell pair between a qubit (E) in the system and the reference qubit, we can first measure $Z_{E}Z_{R}$ and then measure $X_{E}X_{R}$, which is equivalent to apply two-qubit gates $\frac{I \pm Z_{E}Z_{R}}{2}$ and $\frac{I \pm X_{E}X_{R}}{2}$ according to the Born rules.
    
    The von Neumann entropy of subsystem $\alpha$ is $S_{\alpha} = L_{\alpha} - \vert \mathcal{G_{\alpha}} \vert$, where $\mathcal{G}_{\alpha} = \{g_{\alpha} \vert g_{\alpha} \otimes I_{\bar{\alpha}} \in \mathcal{G} \}$ is the canonicalized stabilizer group for reduced density matrix $\rho_{\alpha}$. Therefore, the mutual information between left and right half chains is $I_{A:B} = \vert \mathcal{G}_{AB} \vert - \vert \mathcal{G}_{A} \vert- \vert \mathcal{G}_{B} \vert $ and the mutual information between the system and the reference qubit is $I_{AB:R} = \vert \mathcal{G}_{AB \cup R} \vert - \vert \mathcal{G}_{AB} \vert- \vert \mathcal{G}_{R} \vert $.

\end{document}